\documentclass[showpacs,amsmath,amsart,twocolumn,showkeys,pra,superscriptaddress]{revtex4-1}
\usepackage{dcolumn}    
\usepackage{ifpdf}
\usepackage{amssymb,lineno,amsfonts}
\usepackage{graphicx}   
\usepackage{dcolumn}    
\usepackage{bm}         
\usepackage{bbm}
\usepackage{mathrsfs}
\usepackage{upgreek}
\usepackage{mathtools}
\usepackage{epstopdf}
\usepackage{setspace}
\usepackage{hyperref}
\usepackage{float}
\usepackage{natbib}
\usepackage[usenames,dvipsnames]{xcolor}
\usepackage[matrix,frame,arrow]{xypic}

\newcommand{\ket}[1]{\vert{#1}\rangle}
\newcommand{\bra}[1]{\langle{#1}\vert}
\newcommand{\outpr}[2]{\vert{#1}\rangle\langle{#2}\vert}
\newcommand{\inpr}[2]{\langle{#1}\vert{#2}\rangle}
\newcommand{\expec}[1]{\langle{#1}\rangle}
\definecolor{med-blue}{RGB}{25,25,112}
\hypersetup{colorlinks, linkcolor={red},citecolor={blue}, urlcolor={MidnightBlue}}

\begin{document}

\title{Quantum Cloning using Protective Measurement}
\author{C S Sudheer Kumar} 
\email{sudheer.kumar@students.iiserpune.ac.in}
\affiliation{NMR Research Center, Dept. of Physics, Indian Institute of Science Education and Research, Pune 411008, India} 

\begin{abstract}
Here we show that, in \textit{principle} it is possible to clone (measure) a single arbitrary unknown quantum state of a spin-$\frac{1}{2}$ particle (an electron) with arbitrary precision and with success probability tending to one, using protective measurement. We first transfer the information from spin to spatial degree of freedom (d.o.f) of system electron, then trap it in a double well potential, and finally measure it protectively using a probe electron (which \textit{donot} get entangled with system electron, but still extracts \textit{expectation value} of an observable from a \textit{single} quantum system (system electron)) to obtain information about the unknown spin polarization. Nonorthogonal state discrimination being a subclass of cloning, part of the paper (till finding out $\theta_{m}$, polar angle corresponding to the unknown spin polarization) is sufficient for discrimination.
\end{abstract}

\maketitle
\section{Motivation}
`` There is no law other than the law that there is no law...All is mutable ''$-$ J A Wheeler \cite{Science_and_ulti_realty}. Then how about the `no-cloning' law? Assume there exists a unitary operator $U$ which can clone two non-orthogonal states as follows:
\begin{eqnarray}
U\ket{0}\ket{0}=\ket{0}\ket{0},~U\ket{+}\ket{0}=\ket{+}\ket{+}
\label{no_clone_theorem}
\end{eqnarray}
where, $\ket{+}=(\ket{0}+\ket{1})/\sqrt{2}$, and, $\{\ket{0},\ket{1}\}$ are eigenkets of $z$-component of total spin angular momentum ($S_{z}$). Now taking the inner product of two equatons in \ref{no_clone_theorem}, we get $1=\sqrt{2}$, which is wrong. Hence, generalizing, we can say, it is impossible to clone an arbitrary unknown quantum state (which is non-orthogonal to the reference state $\ket{0}$, in eq. \ref{no_clone_theorem}, of course excluding $\ket{0}$) via a single unitary operation, which is the essence of no-cloning theorem. Now instead consider the following operation:
\begin{eqnarray}
D\ket{0}\ket{0}=\ket{0}\ket{0},~D\ket{+}\ket{0}=\ket{+}\ket{+},\label{clone_via non unitary1} \\
D=\sum\limits_{k,l=0}^{3}h_{kl}\sigma_k\otimes\sigma_l~~~~~~~~~~~~~~\label{D decompsd}
\end{eqnarray}
where, $D$ is an arbitrary linear operator on the Hilbert space, $\mathcal{H}_2\otimes\mathcal{H}_2$, of two spin-1/2 particles. Operators $\{\frac{1}{2}\sigma_k\otimes\sigma_l;k,l=0,1,2,3\}$, where, $\sigma_0=\mathbbm{1}$ ($2\times 2$ identity matrix) and $\sigma_i$'s ($i\ne 0$) are Pauli matrices, form an orthonormal basis in the product Liouville space $\mathbb{L}\otimes\mathbb{L}$ of operators on $\mathcal{H}_2\otimes\mathcal{H}_2$ \cite{Audruch_entangled_sys}. $h_{kl}$ are the coefficients of decomposition of $D$ in the basis $\{\frac{1}{2}\sigma_k\otimes\sigma_l\}$. Therefore, we have $16$ unknown complex coefficients $h_{kl}$ i.e., $32$ real coefficients and $16$ constraint equations (8 real $+$ 8 imaginary, obtained from eq.s in \ref{clone_via non unitary1}). Hence we can express $16$ real coefficients in terms of $16$ other arbitrary real parameters. This shows \textit{there exists infinitely many solutions to eqations in} \ref{clone_via non unitary1} (for one such solution see \footnote[4]{\begin{eqnarray}
D=\frac{1}{\sqrt{2}}
  \begin{bmatrix}
    \sqrt{2}&~0  & ~1-\sqrt{2} &~ 0 \\0&~0  &~~~1 &~ 0\\0&~0  &~~~1 &~ 0\\0&~0  &~~~1 &~ 0        \end{bmatrix} 
    \label{D nonuni op a sp eg}
\end{eqnarray} One can easily verify that this $D$ satisfies equations in \ref{clone_via non unitary1}, and it is possible to decompose it as in eq. \ref{D decompsd}.}). $D$ being sum of unitary operators, $\sigma_k\otimes\sigma_l$, is a non-unitary operator in general and specifically in eq. \ref{clone_via non unitary1}. It may not be possible to \textit{directly} implement the operator $D$ in an experiment (however, there might be some indirect way). As there are infinitely many solutions to eq.s in \ref{clone_via non unitary1}, many of them may be just mathematical objects with no relevance to physical operations. Now let us generalize the case in eq. \ref{clone_via non unitary1}:
\begin{eqnarray}
D\ket{0}\ket{0}=\ket{0}\ket{0},~D\ket{\hat{m}}\ket{0}=\ket{\hat{m}}\ket{\hat{m}},\label{clone_via non unitary1_gen} \\
D=\sum\limits_{k,l=0}^{3}h_{kl}\sigma_k\otimes\sigma_l~~~~~~~~~~~~~~\label{D decompsd_gen}
\end{eqnarray}
where, $\ket{\hat{m}}$ (given by eq. \ref{unknown state}) is the unknown state to be cloned. Again there exists infinitely many solutions to eq.s in \ref{clone_via non unitary1_gen} (for one such solution see \footnote[5]{\begin{eqnarray}
D=
  \begin{bmatrix}
    1&~0  & ~e^{-i\phi_m}\cot\frac{\theta_m}{2}(\cos\frac{\theta_m}{2}-1) &~ 0 \\0&~0  &~~~\cos\frac{\theta_m}{2} &~ 0\\0&~0  &~~~\cos\frac{\theta_m}{2} &~ 0\\0&~0  &~~~e^{i\phi_m}\sin\frac{\theta_m}{2} &~ 0        \end{bmatrix} 
    \label{D nonuni op a gen eg}
\end{eqnarray} It is not possible to know the operator $D$ apriori, as it is a function of unknown parameters $\theta_m$ and $\phi_m$. In the protocol that we are going to describe, we will come to know $\theta_m$ and $\phi_m$, only at the end. The whole chain of complex processes involved in the protocol, may not be representable by a simple operator like $D$. Here we are just trying to motivate mathematically and $D$ may not have any relevance to physical reality at all! }). In eq. \ref{clone_via non unitary1_gen}, if $\ket{\hat{m}}=\ket{1}$ then, $D\ket{1}\ket{0}=\ket{1}\ket{1}$. But, $D$ is a \textit{linear} operator. Hence, 
\begin{eqnarray}
D\ket{\hat{m}}\ket{0}=\cos\frac{\theta_{m}}{2}D\ket{0}\ket{0}+\sin\frac{\theta_{m}}{2}e^{i\phi_{m}}D\ket{1}\ket{0}\label{linearity imposed}\\
=\cos\frac{\theta_{m}}{2}\ket{0}\ket{0}+\sin\frac{\theta_{m}}{2}e^{i\phi_{m}}\ket{1}\ket{1}\ne\ket{\hat{m}}\ket{\hat{m}}
\label{linerity noclone}
\end{eqnarray}
Hence, if $D$ is linear, we cannot clone an arbitrary unknown state \cite{noclone_review}. Hence, $D$ must be a \textit{nonlinear operator} (i.e., eq. \ref{linearity imposed} doesnot hold) apart from being \textit{nonunitary}, to clone an arbitrary unknown quantum state. The mysterious \textit{quantum measurement} process, which involves amplifying information to the classical limit, is one such operation which is both nonlinear \footnote[8]{Let $Q$ be a macroscopic measuring device with very large number of d.o.f (as it amplifies the information) which measures $S_z$ ($z$-component of total spin angular momentum operator). Then we have: $Q|0>=\frac{\hbar}{2}|0>$ and $Q|1>=\frac{-\hbar}{2}|1>$. Now consider: 
\begin{eqnarray}
Q(|0>+|1>)/\sqrt{2}=
\begin{cases}
    \frac{\hbar}{2}\frac{1}{\sqrt{2}}|0>,~\mathrm{with~probability}~1/2\\
    ~~~~~~~~~~~~~\mathrm{OR}\\
    \frac{-\hbar}{2}\frac{1}{\sqrt{2}}|1>,~\mathrm{with~probability}~1/2\\
\end{cases} \\
\ne (Q|0>+Q|1>)/\sqrt{2}=\frac{\hbar}{2}(|0>-|1>)/\sqrt{2}~~~~~~
\label{mesurmnt nonlinear} 
\end{eqnarray} Hence measurement operator $Q$, is a nonlinear operator, as it breaks the superposition. In the protocol that we are going to describe, probe electron starts with $\Delta p_{az}\rightarrow 0$ (footnote below eq. \ref{p_aoz,final}) and it will not get entangled with system electron, but still gains momentum proportional to $\theta_m$. As explained in footnote below eq. \ref{p_aoz,final}, in principle it is possible to measure this momentum (and hence $\theta_m$) with arbitrary precision. This can be represented as: $\tilde{Q}\bar{\Theta}\rightarrow p_{a0z}^{final}\bar{\Theta}$ (this is similar to $Q|0>=\frac{\hbar}{2}|0>$), where, $\tilde{Q}$ is a macroscopic measuring device similar to $Q$, and $\bar{\Theta}$ is the state of probe electron. This shows there is no collapse (as no entanglement any where) upon measuring probe electron and hence we are able to get $p_{a0z}^{final}$ with probability $\rightarrow$ one. This does not mean that we have obtained the value of $p_{a0z}^{final}$ with a linear operation (as no collapse has occured), because $\tilde{Q}$ is intrinsically a nonlinear operator. Its nonlinearity has not been explicitly displayed, because our protocol was such that it avoided entanglement all the way. It is similar to the following situation: Let $f(x)=x^2$ $\Rightarrow$ $f(x+0)=f(x)+f(0)$. This does not mean that $f$ is a linear function. We have chosen the values in such a way that the function \textit{appears} to behave like a linear function. Now, $p_{a0z}^{final}$ has to be solved for $\theta_m$ using an equation similar to \ref{p_aoz,final}. This process of solving (which is very much part of cloning) is a macroscopic phenomenon (measurement) which involves directly or indirectly collapse of wave function (eg., as we write down the equation on a piece of paper, we are collapsing the state of atoms of ink, which otherwise can exist anywhere in the universe with nonzero probability, however small), there by nonlinear processes explicitly come into picture. There are still many such processes which directly or indirectly involve nonlinear processes (which are also very much part of cloning) like, solving constraint equations \ref{cos theta mn evalted}, \ref{cos theta ml evalted}, and, clubbibg the information $\theta_m,~\phi_m$ and preparing another qubit in the state $|\hat{m}>$. We also note that, according to Everett's many-worlds interpretation and Bohm's causal interpretation of quantum measurement \cite{D_Home_book}, there is no breaking of superposition at the level of consciousness and matter respectively, and hence no nonlinearity. However, we believe that, as we cannot observe superposed states, some where nonlinearity has to come in, either at the level of matter or consciousness.} \cite{D_Home_book} and nonunitary (as information is irreversibly amplified to the classical limit which increases entropy, eg., electron absorbed by the screen in Stern-Gerlach experiment). Hence, it motivates us to ask the following question: `Is it possible to clone an arbitrary unknown quantum state, through a combination of linear unitary and nonlinear nonunitary (measurement) operations ?'. We found the answer to be `Yes'. In the protocol that we are going to describe, we carry out many such measurements (nonlinear nonunitary) eg., on probe electrons (which donot get entangled with system electron, but still extracts information from it!), there by \textit{relaxing both unitarity and linearity} constraints, which allows us to clone. In the following protocol, even though operations on system electron are linear unitary, operations on probe electrons and other auxiliary systems are nonlinear nonunitary, and clubbing informations obtained via measurements (which is some kind of sum of unitaries which is nonunitary in general) to prepare another qubit in the state same as that of given qubit, and to get a final state similar to that on RHS of eq. \ref{clone_via non unitary1_gen}, makes the global (system+probes+observer) process nonlinear nonunitary.\\Finally we like to mention that, in the cloning protocol that we are going to describe, in principle we can clone a single arbitrary unknown quantum state of a spin-$1/2$ particle (electron) with success probability tending to one (but not exactly equal to one, and hence there is probability, however small, of failure) and with arbitrary precision (which, strictly speaking, does not correspond to an exact or identical copy, as there is an error, however small). Hence, strictly speaking, it is not perfect cloning (exactly zero error) with success probability exactly equal to one. 

\section{Cloning Protocol}
\begin{figure}[H]
\centering
\includegraphics[width=7.5cm]{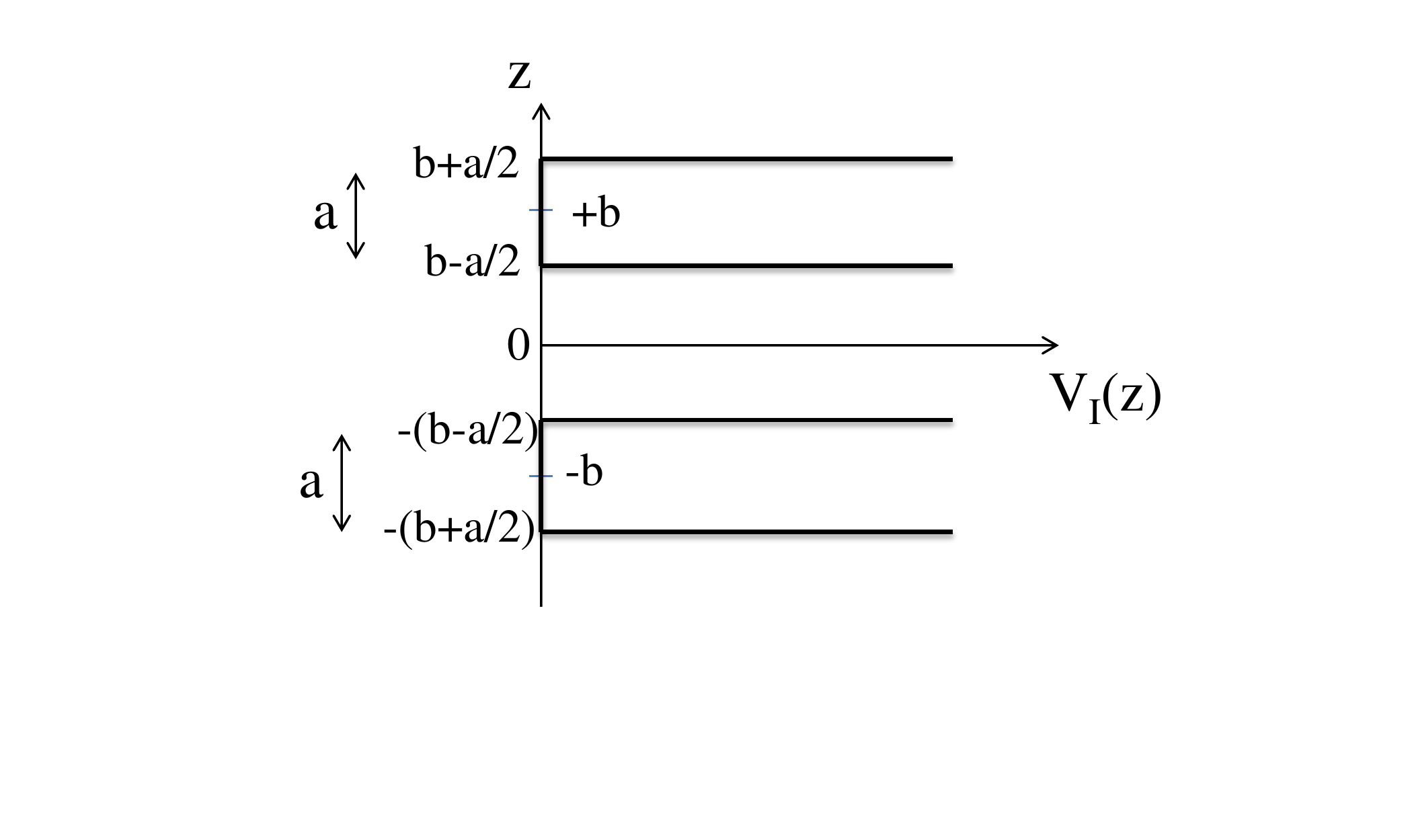}
\caption{Symmetrical infinite double well potential} 
\label{fig_double_well}
\end{figure}

Protective Measurement was invented by Y Aharonov and L Vaidman in their seminal paper \cite{prot_measr_birth_of}. The cloning protocol that we are going to describe is motivated by an idea proposed by Y Aharonov, J Anandan and L Vaidman in \cite{prot_measure}. Let the system be an electron which is a spin-$\frac{1}{2}$ particle with charge $-e$ and mass $M$. Let its spin be in an arbitrary unknown state: 
\begin{eqnarray}
\ket{\hat{m}}=\cos\frac{\theta_{m}}{2}\ket{0}+\sin\frac{\theta_{m}}{2}e^{i\phi_{m}}\ket{1}
\label{unknown state}
\end{eqnarray} 
where $\hat{m}$ is a unit vector along unknown spin polarization given by,
\begin{equation}
\hat{m}=\sin\theta_{m}\cos\phi_{m}\hat{i}+\sin\theta_{m}\sin\phi_{m}\hat{j}+\cos\theta_{m}\hat{k}
\label{m cap decomposed}
\end{equation}
where, $\hat{i},\hat{j},\hat{k}$ are unit vectors along positive $x,y,z$ axes respectively of fixed lab frame, $\theta_m$ and $\phi_m$ are polar and azimuthal angles respectively. For convenience, let the system electron's wave packet be a Gaussian centered at $z_{0}=0,~p_{0z}$(average momentum)$=0$, which satisfies $\Delta z \Delta p_{z}=\hbar/2$, say, at time $t=0$. $\Delta z$ and $\Delta p_{z}$ are standard deviations in position and momentum respectively. The variance $\Delta p_{z}^{2}$, can be Squeezed (at the cost of increasing $\Delta z^{2}$) and made small enough (but not zero, because, if zero then we have a plane wave, which is not square integrable and hence, rigorously, cannot represent a physical state of system electron (p23 \cite{QM_Cohen})) to be taken as zero for all practical purposes. For eg., if $\Delta z$ is of the order of $10^{-3}m$ then $\Delta p_{z}$ will be of the order of $10^{-31}kg~m~s^{-1}$, which can be treated as zero in conjunction with eq. \ref{Delta E}, for all practical purposes. Also, in principle we can make $\Delta p_{z}$ arbitrarily close to zero (but not exactly zero, there by satisfying square integrability requirement). To compensate for the corresponding increase in $\Delta z$, in principle we can make width `$a$' of potential well (Fig.\ref{fig_double_well}), arbitrarily large as explained below eq. \ref{ket phi_dwp}. Hence, in principle we can push the system electron into one of the eigenstates of $H_{0S}$(\ref{H_0S}) or $H_{S}$ (\ref{H_S defined}), as discussed below eq.s \ref{step f'n} and \ref{enrgy cr.ing variance} respectively. \\As the wave packet evolves, it spreads in position space but not in momentum space. Its average momentum ($p_{0z}$) as well as momentum dispersion $\Delta p_z$, do not change with time. Hence, at $t>0$, $\Delta z \Delta p_{z}>\hbar/2$. Because $p_{0z}=0$, even $z_0$ (center of wave packet in position space) do not change with time (pp 64-65 of \cite{QM_Cohen}).  Now an inhomogeneous magnetic field along $z$-axis is switched on for an interval of time $\tau$. System electron evolves under the total Hamiltonian $p_z^2/(2M)+H_{int}$, where, interaction Hamiltonian, $H_{int}=-\vec{\mu}\cdot B_{i}z\hat{k}=-\gamma B_{i}S_{z}z$, $B_{i}$ is the constant gradient($Tesla~m^{-1}$) in magnetic field along $z$-axis \footnote[1]{In an actual Stern Gerlach apparatus both $\vec{\nabla}\cdot\vec{B}=0$ and $\vec{\nabla}\times\vec{B}=0$. The field that satisfies these conditions is: $\vec{B}=B\hat{B}=-B_ix~\hat{i}+(B_0+B_iz)\hat{k}$ (\cite{stern_gerlach_analysed}, p388 of \cite{QM_Cohen}), where, $B_0>0$. For $B_0>>|B_i|$, $\hat{B}\approxeq\hat{k}$ and hence spin precesses about $z$-axis with frequency $\omega\approxeq-\gamma B_0$. As a result, $x$-component of spin, $S_x$, oscillates in time and hence its average over an interval of time large compared to period of oscillation ($2\pi/\omega$) but small compared to $\tau$, vanishes. Only $S_z$ survives. Hence wave packet splits only along $z$-axis. Due to the precence of static field $B_0\hat{k}$, unknown spin precesses about $z$-axis, which can be easily taken into account. Hence we are not going to consider the effect of static field.}, $\gamma$ is the gyromagnetic ratio of electron which is negative. As $p_{0z}=0$, in the limit $\Delta p_z\rightarrow 0$ (case of our interest, see below eq. \ref{ket phi_dwp}), $p_z^2/(2M)\rightarrow 0$. However, as it evolves under $H_{int}$, it gains kinetic energy. We can choose static field $B_0$ \footnote[1]{} such that actual interaction energy  $H'_{int}>>p_z^2/(2M)$ for any $B_i$, $\tau$ \footnote[15]{From the previous footnote it is evident that actual interaction energy, $H'_{int}=-\vec{\mu}\cdot\vec{B}=-\gamma B_0 S_z-\gamma B_i S_z z+\gamma B_i S_x x $. Eigenvalues os $S_z$ are $\pm\hbar/2$. As the reference point can always be shifted, we consider only the difference in energy levels which is: $-\gamma B_0\hbar>0$ as $\gamma<0$. Hence we can write: $H'_{int}= |\gamma| B_0 \hbar-|\gamma B_i|\hbar z+|\gamma B_i| \hbar x$. As $B_0$ can be arbitrarily large, $H'_{int}$ can also be made arbitrarily large for any $B_i,z,x$. Hence, $H'_{int}>>p_z^2/(2M)$ holds for any $B_i$, $\tau$ }. Hence, we can neglect the evolution under $p_z^2/(2M)$, and write:
\begin{eqnarray}
\ket{\psi(\tau,p_{z})}=\exp(-\frac{i}{\hbar}H_{int}\tau)\ket{\hat{m}}\bar{\chi}(p_{z}-0)=~~~~~~~~~~~~~~~\nonumber\\
\cos\frac{\theta_{m}}{2}\ket{0}\bar{\chi}(p_{z}-\gamma B_{i}\frac{\hbar}{2}\tau)+\sin\frac{\theta_{m}}{2}e^{i\phi_{m}}\ket{1}\bar{\chi}(p_{z}+\gamma B_{i}\frac{\hbar}{2}\tau)\nonumber\\
\label{splitting stage}
\end{eqnarray}
where $\bar{\chi}(p_{z}-0)$ is the wave packet of system electron in momentum space centered at $p_{0z}=0$ (see eq. \ref{genrtr of momentum} for definition of wave packet), which has split and got entangled with its spin d.o.f. Taking its inverse Fourier transform and finding its time evolution, we see that they are still entangled and the split packets are moving in opposite directions. Gradient $B_i$ can be made either negative (p388 of \cite{QM_Cohen}) or positive. Let $B_i<0$. Then, system electron is in a superposition state of having momentum $\gamma B_{i}\frac{\hbar}{2}\tau$ (hence moving along positive $z$-axis) with probability $\cos^{2}\frac{\theta_{m}}{2}$ and having momentum $-\gamma B_{i}\frac{\hbar}{2}\tau$ (hence moving along negative $z$-axis) with probability $\sin^{2}\frac{\theta_{m}}{2}$. Let's trap the system electron in a symmetrical double well potential (trapping procedure explained below eq. \ref{ket phi_dwp}) with the following description(see for eg.,\cite{QM_Cohen,empty_wave_hardy}),
\begin{eqnarray}
V_{I}(z)=
\begin{cases}
    0,~\mathrm{for}~(b-a/2)<z<(b+a/2)\\
    0,~\mathrm{for}~-(b+a/2)<z<-(b-a/2)\\
    \infty,~\mathrm{everywhere~else}\\
\end{cases} 
\label{box potential defined}
\end{eqnarray}
where, $a$ is the length of each potential well and $b$ is the distance from origin of coordinate system to center of potential well(see Fig.\ref{fig_double_well}). $V_I(z)$ is the potential experienced by system electron \textit{only}, but not by any other charges outside the potential well.

System electron has gained an average momentum $\gamma B_{i}\frac{\hbar}{2}\tau$ and hence its average kinetic energy is $\frac{p_{z}^{2}}{2M}=\frac{1}{2M}(\gamma B_{i}\frac{\hbar}{2}\tau)^2$, where all the parameters are known. We can choose the values of free parameters $a,b,B_i$ and $\tau$ (as explained below eq. \ref{step f'n}), such that the \textit{system electron enters one of the eigenstates of the system Hamiltonian}:
\begin{equation}
H_{0S}=\dfrac{p_{z}^{2}}{2M}+V_{I}(z)
\label{H_0S}
\end{equation} 
where, $V_{I}(z)$ is as given in eq. \ref{box potential defined}. State of the system electron after trapping it in double well potential is given by,
\begin{widetext}
\begin{eqnarray}
\ket{\Phi_{0sn}(z)}=\cos\frac{\theta_{m}}{2}\ket{0}\left[ \theta(z-(b-\frac{a}{2}))-\theta(z-(b+\frac{a}{2}))\right] \sqrt{\frac{2}{a}}\sin\big(k_{0sn}(b+\frac{a}{2}-z)\big)\nonumber\\
+\sin\frac{\theta_{m}}{2}e^{i\phi_{m}}\ket{1}\left[ \theta(z+(b+\frac{a}{2}))-\theta(z+(b-\frac{a}{2}))\right] \sqrt{\frac{2}{a}}\sin\big(k_{0sn}(b+\frac{a}{2}+z)\big)
\label{state inside infinite well}
\end{eqnarray}
\end{widetext}
where, $k_{0sn}=\frac{n\pi}{a},~n=1,2,3,...$. Unit step function is defined as,
\begin{eqnarray}
\theta(z-z_{0})=
\begin{cases}
    0,~\mathrm{for}~z<z_{0}\\
    1,~\mathrm{for}~z>z_{0}\\
\end{cases}
\label{step f'n} 
\end{eqnarray}
which implies $\frac{d}{dz}\theta(z-z_{0})=\delta(z-z_{0})$, Dirac-Delta function. We note that, even though $H_{0S}$ commutes with parity operator, its eigenstate $\ket{\Phi_{0sn}(z)}$ is not symmetric due to entanglement with spin d.o.f. Similarly $\ket{\Phi_{0an}(z)}$ (\ref{anti sym state inside infinite well}) is not antisymmetric. Eq. \ref{state inside infinite well} describes the superposition state of system electron being in upper and lower potential wells, while entangled with its spin degree of freedom. When $\theta_{m}=0$ or $\pi$, there is no splitting of the wave packet and eq. \ref{state inside infinite well} reduces to the correct form describing the unsplit wave packet inside the potential well. 

Inside the double well potential (excluding the boundary points, as derivative of the wave function does not exist there) $H_{0S}=\dfrac{p_{z}^{2}}{2M}=-\frac{\hbar^{2}}{2M}\frac{d^{2}}{dz^{2}}$ and it is easy to check, using $\frac{d}{dz}\theta(z-z_{0})=\delta(z-z_{0})$, that the state $\ket{\Phi_{osn}(z)}$ given in eq.\ref{state inside infinite well} is an eigenket of $H_{0S}$ with eigenvalue $E_{n}=\frac{\hbar^{2}k_{0sn}^{2}}{2M}$. 

We can make the system electron to enter the eigenstate $\ket{\Phi_{0sn}(z)}$ of $H_{0S}$, by choosing $E_{n}=\frac{p_{z}^{2}}{2M}=\frac{1}{2M}(\gamma B_{i}\frac{\hbar}{2}\tau)^2$ and following an argument similar to that carried out from eq. \ref{enrgy cr.ing variance} to \ref{ket phi_dwp} with the limit $V_0\rightarrow\infty$ (here we also take the limit $\Delta p_{z}\rightarrow 0$. However, this is not necessary if we follow a different technique as explained in para next to next  to eq. \ref{ket phi_dwp}). Hence, in principle we can push the system electron into one of the eigenstates of $H_{0S}$. \\Wavefunction in eq. \ref{state inside infinite well} is \textit{Square Integrable} unlike the plane waves  $\exp(\frac{i}{\hbar}p_{z}z)$ and hence represents the actual physical state of system electron. Hence there is no need to construct a wave packet using the eigenkets $\{\ket{\Phi_{0sn}(z)}\}$, to represent the physical state of system electron inside the potential well, unlike the situation when the electron was free. This justifies our assumption that system electron enters the eigenstate given in eq. \ref{state inside infinite well}, inside the potential well. This can be further justified by the fact that, an electron in a bound state of hydrogen atom can exist in one of the eigenstates, even though it was in a superposition of infinite number of plane waves $\exp(\frac{i}{\hbar}\vec{p}.\vec{r})$ when it was unbounded. 

\textbf{Completeness of eigenstates of $H_{0S}$:} Using eq.\ref{state inside infinite well}, taking the inner product of $\ket{\Phi_{0sk}(z)}$ with $\ket{\Phi_{0sn}(z)}$ which involves integration w.r.t $z$ from $-\infty$ to $+\infty$, we get $\inpr{\Phi_{0sk}(z)}{\Phi_{0sn}(z)}=\delta_{kn}$, where $\delta_{kn}$ is the kronecker delta function, and hence orthonormal. In eq. \ref{state inside infinite well} if we introduce a relative phase ($e^{i\pi}$) between two terms (which is equivalent to changing $\theta_m$ to $-\theta_m$), we get a normalized solution which is linearly independent from $\ket{\Phi_{0sn}(z)}$. Then using Grahm-Schmidt orthogonalization, the solution becomes:
\begin{widetext}
\begin{eqnarray}
\ket{\Phi_{0an}(z)}=\sin\frac{\theta_{m}}{2}\ket{0}\left[ \theta(z-(b-\frac{a}{2}))-\theta(z-(b+\frac{a}{2}))\right] \sqrt{\frac{2}{a}}\sin\big(k_{0an}(b+\frac{a}{2}-z)\big)\nonumber\\
-\cos\frac{\theta_{m}}{2}e^{i\phi_{m}}\ket{1}\left[ \theta(z+(b+\frac{a}{2}))-\theta(z+(b-\frac{a}{2}))\right] \sqrt{\frac{2}{a}}\sin\big(k_{0an}(b+\frac{a}{2}+z)\big)
\label{anti sym state inside infinite well}
\end{eqnarray}
\end{widetext}
where $k_{0an}=\frac{n\pi}{a},~n=1,2,3,...$. One can verify that $\ket{\Phi_{0an}(z)}$ is an eigenket of $H_{0S}$, with eigenvalue same as that corresponding to the state $\ket{\Phi_{0sn}(z)}$, and hence degenerate with it. One can also verify that $\inpr{\Phi_{0ak}(z)}{\Phi_{0an}(z)}=\delta_{kn}$ and $\inpr{\Phi_{0sk}(z)}{\Phi_{0an}(z)}=0$ for all $k,n$ and hence mutually orthogonal. Using the relation:
\begin{widetext}
\begin{eqnarray}
\sum\limits_{n=1}^{\infty}\sqrt{\frac{2}{a}}\sin\big(\frac{n\pi}{a}z\big)\sqrt{\frac{2}{a}}\sin\big(\frac{n\pi}{a}z'\big)\left[ \theta(z)-\theta(z-a)\right]\left[ \theta(z')-\theta(z'-a)\right]=\delta(z'-z)\left[ \theta(z)-\theta(z-a)\right]\left[ \theta(z')-\theta(z'-a)\right]\nonumber\\
\label{delta fn}
\end{eqnarray}
\end{widetext}
where, $\delta(z'-z)$ is the Dirac delta function, one can verify that:
\begin{widetext}
\begin{eqnarray}
\sum\limits_{n=1}^{\infty}\bigg(\outpr{\Phi_{0sn}(z)}{\Phi_{0sn}(z')}+\outpr{\Phi_{0an}(z)}{\Phi_{0an}(z')}\bigg)=\outpr{0}{0}\delta(z'-z)\left[ \theta(z-(b-\frac{a}{2}))-\theta(z-(b+\frac{a}{2}))\right]\nonumber\\\left[ \theta(z'-(b-\frac{a}{2}))-\theta(z'-(b+\frac{a}{2}))\right]+\outpr{1}{1}\delta(z'-z)\left[ \theta(z+(b+\frac{a}{2}))-\theta(z+(b-\frac{a}{2}))\right]\nonumber\\\left[ \theta(z'+(b+\frac{a}{2}))-\theta(z'+(b-\frac{a}{2}))\right]~~~~~~\Rightarrow\int\limits_{-\infty}^{\infty}dz'\sum\limits_{n=1}^{\infty}\bigg(\outpr{\Phi_{0sn}(z)}{\Phi_{0sn}(z')}+\outpr{\Phi_{0an}(z)}{\Phi_{0an}(z')}\bigg)=\mathbbm{1}
\label{compltness rel}
\end{eqnarray}
\end{widetext}
where, $\mathbbm{1}$ is $2\times 2$ identity matrix, there by satisfying the completeness relation.

To measure a given system protectively, system should be in a non-degenerate eigenstate of system Hamiltonian. But the state $\ket{\Phi_{0sn}(z)}$, which we want to measure protectively, is a degenerate eigenstate of $H_{0S}$, as discussed in previous paragraph. If we perturb the system by applying a small negative potential in the region $-(b-\frac{a}{2})\le z\le(b-\frac{a}{2})$, degeneracy will be lifted due to tunneling. Adding a small negative potential to a very large but finite potential(which is infinite for all practical purposes), lowers the net potential slightly, there by allowing tunneling. Instead of solving perturbatively (for perturbative treatment see \cite{QM_Cohen}), we are going to treat the potential in the region $-(b-\frac{a}{2})\le z\le(b-\frac{a}{2})$ to be $V_{0}(>0)$, which is finite and constant, and solve exactly. If necessary, later we can take $V_{0}$ to be very large but finite, to satisfy the condition of small perturbation to $V_{I}(z)$ (\ref{box potential defined}). Following treatment is similar to that on pp460-62 of \cite{QM_Cohen}. We are going to find the eigenstates $\ket{\phi_{sn}(z)}$, $\ket{\phi_{an}(z)}$ of new system Hamiltonian:
\begin{equation}
H_{S}=\dfrac{p_{z}^{2}}{2M}+V_{F}(z)
\label{H_S defined}
\end{equation}
with eigenvalues $E_{sn}$ and $E_{an}$ respectively, which are less than $V_{0}$. $V_{F}(z)$ is same as $V_{I}(z)$ given in eq. \ref{box potential defined} but $V_{F}(z)=V_{0}$ in the region $-(b-\frac{a}{2})\le z\le(b-\frac{a}{2})$ instead of $\infty$.

As the system electron is in superposition of being in both wells, it can tunnel from both wells into the region $-(b-a/2)<z<(b-a/2)$. As the split wave packets tunnel into the region $-(b-a/2)<z<(b-a/2)$, they get united and hence disentangles from spin d.o.f. This is justified by the fact that, if we reverse the direction of inhomogeneous magnetic field (see above eq. \ref{n cap defined}), the split wave packets start moving towards the origin, get united and hence disentangles from spin d.o.f. With this requirement, $\ket{\Phi_{0sn}(z)}$ (\ref{state inside infinite well}) suggests the following form for one of the eigenstates of $H_{S}$ with energy $E_s<V_0$:
\begin{widetext}
\begin{eqnarray}
\ket{\phi_s(z)}=\cos\frac{\theta_{m}}{2}\ket{0}\left[ \theta(z-(b-\frac{a}{2}))-\theta(z-(b+\frac{a}{2}))\right] A~\sin\big(k_s(b+\frac{a}{2}-z)\big)\nonumber\\
+\sin\frac{\theta_{m}}{2}e^{i\phi_{m}}\ket{1}\left[ \theta(z+(b+\frac{a}{2}))-\theta(z+(b-\frac{a}{2}))\right] A'~\sin\big(k_s(b+\frac{a}{2}+z)\big)\nonumber\\+(\cos\frac{\theta_{m}}{2}\ket{0}+\sin\frac{\theta_{m}}{2}e^{i\phi_{m}}\ket{1})
\left[ \theta(z+(b-\frac{a}{2}))-\theta(z-(b-\frac{a}{2}))\right](B~e^{q_s(z-b)}+B'~e^{-q_s(z-(-b))})
\label{gen state finite well}
\end{eqnarray}
\end{widetext}
where, $q_s=\sqrt{\frac{2M}{\hbar^{2}}(V_{0}-E_s)}=\sqrt{\alpha^{2}-k_s^{2}}~>0$. We have related energy $E_s$ of the state with its wave number $k_s$ by the relation:
\begin{equation}
E_s=\frac{\hbar^{2}k_s^{2}}{2M}
\label{E reltd to k}
\end{equation}
Exponentially decaying wave $e^{q_s(z-b)}$ corresponds to tunnelling from upper potential well (i.e., potential well on positive $z$-axis) into the region $-(b-\frac{a}{2})\le z\le(b-\frac{a}{2})$ and $e^{-q_s(z-(-b))}$ corresponds to that from lower well. Further, $\ket{\Phi_{0sn}(z)}$ (\ref{state inside infinite well}) suggests to take $A=A'=A_{s}$ and $B=B'=B'_{s}$ in eq. \ref{gen state finite well}. Evanescent wave reduces to the form: $B~e^{q_{s}(z-b)}+B'~e^{-q_{s}(z-(-b))}=B_{s}~\cosh(q_{s}z)$, where $B_{s}=2B'_{s}e^{-q_{s}b}$. It is evident that in the limit $V_{0}$ (hence $q_s$) going to infinity, both evanescent waves vanish and we recover the state in eq. \ref{state inside infinite well} as required. Demanding that the spatial wave function and its first derivative must be continuous at $z=b-\frac{a}{2}$, we obtain the following constraint equation:
\begin{eqnarray}
\tan(k_{s}a)=-\dfrac{k_{s}}{\sqrt{\alpha^{2}-k^{2}_{s}}}\coth\big(\sqrt{\alpha^{2}-k^{2}_{s}}(b-\frac{a}{2})\big)~~~~~~~~~
\label{constraint eq sym}
\end{eqnarray}
This follows from the fact that $A_{s}$ and $B_{s}$ cannot vanish simultaneously, else we get trivial solution $\ket{\phi_{s}(z)}=0$ everywhere. Also in obtaining eq. \ref{constraint eq sym} we dropped unit step function which is justifiable as it is used just for the sake of convenience. Similar joining conditions at $z=-(b-\frac{a}{2})$, also gives same constraint eq. \ref{constraint eq sym}.\\
Now, $\ket{\Phi_{0an}(z)}$ (\ref{anti sym state inside infinite well}) suggests the following form for another eigenstate of $H_{S}$ (which is linearly independent from $\ket{\phi_s(z)}$ \ref{gen state finite well}) with energy $E_a<V_0$:
\begin{widetext}
\begin{eqnarray}
\ket{\phi_a(z)}=\sin\frac{\theta_{m}}{2}\ket{0}\left[ \theta(z-(b-\frac{a}{2}))-\theta(z-(b+\frac{a}{2}))\right] A~\sin\big(k_a(b+\frac{a}{2}-z)\big)\nonumber\\
+\cos\frac{\theta_{m}}{2}e^{i\phi_{m}}\ket{1}\left[ \theta(z+(b+\frac{a}{2}))-\theta(z+(b-\frac{a}{2}))\right] A'~\sin\big(k_a(b+\frac{a}{2}+z)\big)\nonumber\\+(\sin\frac{\theta_{m}}{2}\ket{0}+\cos\frac{\theta_{m}}{2}e^{i\phi_{m}}\ket{1})
\left[ \theta(z+(b-\frac{a}{2}))-\theta(z-(b-\frac{a}{2}))\right](B~e^{q_a(z-b)}+B'~e^{-q_a(z-(-b))})
\label{gen state finite well anti sym}
\end{eqnarray}
\end{widetext}
where, $q_a=\sqrt{\frac{2M}{\hbar^{2}}(V_{0}-E_a)}=\sqrt{\alpha^{2}-k_a^{2}}~>0$, $E_a=\hbar^2k_a^2/(2M)$. $\ket{\Phi_{0an}(z)}$ (\ref{anti sym state inside infinite well}) suggests to take $A'=-A=-A_{a}$ and $B'=-B=-B'_{a}$. With this, evanescent wave takes the form: $B~e^{q_{a}(z-b)}+B'~e^{-q_{a}(z-(-b))}=B_{a}~\sinh(q_{a}z)$, where, $B_{a}=2B'_{a}e^{-q_{a}b}$. In the limit $V_0\rightarrow\infty$, $\ket{\phi_a(z)}$ (\ref{gen state finite well anti sym}) reduces to $\ket{\Phi_{0an}(z)}$ (\ref{anti sym state inside infinite well}) as required. Applying joining conditions at $z= b-\frac{a}{2}$ in a manner similar to previous case, we obtain the following constraint equation:
\begin{eqnarray}
\tan(k_{a}a)=-\dfrac{k_{a}}{\sqrt{\alpha^{2}-k^{2}_{a}}}\tanh\big(\sqrt{\alpha^{2}-k^{2}_{a}}(b-\frac{a}{2})\big)~~~~~~~~~
\label{constraint eq antisym}
\end{eqnarray}
Joining conditions at $z=-(b-\frac{a}{2})$ also leads to same constraint eq. \ref{constraint eq antisym}. Equtions \ref{constraint eq sym} and \ref{constraint eq antisym} can be solved graphically for $k_{s}$ and $k_{a}$ respectively, to obtain $k_{sn}$ and $k_{an}$ as $n^{th}$ roots, there by quantizing the energy levels. As constraint eq.s \ref{constraint eq sym} and \ref{constraint eq antisym} are not identical in form, $k_{sn}$ will not be equal to $k_{an}$. Then, using relation \ref{E reltd to k} and $E_a=\hbar^2k_a^2/(2M)$, we obtain $E_{sn}\ne E_{an}$, \textit{there by lifting degeneracy}. In the limit $V_{0}\rightarrow\infty$, we obtain from eq.s \ref{constraint eq sym} and \ref{constraint eq antisym}: $k_{s,a}\rightarrow\frac{n\pi}{a}$ as required. As $q_{sn,an}>0$ we have $E_{sn,an}<V_{0}$. Eigenstates of $H_{S}$, given by eqs. \ref{gen state finite well} and \ref{gen state finite well anti sym}, take the following form after quantization:
\begin{eqnarray}
\ket{\phi_{sn}(z)}=\ket{\phi_{0sn}(z)}+\ket{\hat{m}}B_{sn}~\cosh(q_{sn}z)~~~~~~~~~\nonumber\\\left[ \theta(z+(b-\frac{a}{2}))-\theta(z-(b-\frac{a}{2}))\right],~~~~~~\nonumber\\
\ket{\phi_{an}(z)}=\ket{\phi_{0an}(z)}+(\sin\frac{\theta_{m}}{2}\ket{0}+\cos\frac{\theta_{m}}{2}e^{i\phi_{m}}\ket{1})~~~~\nonumber\\B_{an}~\sinh(q_{an}z)\left[ \theta(z+(b-\frac{a}{2}))-\theta(z-(b-\frac{a}{2}))\right]~~~~~~~
\label{state inside finite well}
\end{eqnarray}
where, $\ket{\hat{m}}$ is given by eq. \ref{unknown state}, and:
\begin{widetext}
\begin{eqnarray}
\ket{\phi_{0sn}(z)}=\cos\frac{\theta_{m}}{2}\ket{0}\left[ \theta(z-(b-\frac{a}{2}))-\theta(z-(b+\frac{a}{2}))\right] A_{sn}~ \sin\big(k_{sn}(b+\frac{a}{2}-z)\big)\nonumber\\
+\sin\frac{\theta_{m}}{2}e^{i\phi_{m}}\ket{1}\left[ \theta(z+(b+\frac{a}{2}))-\theta(z+(b-\frac{a}{2}))\right] A_{sn}~\sin\big(k_{sn}(b+\frac{a}{2}+z)\big),\nonumber\\
\ket{\phi_{0an}(z)}=\sin\frac{\theta_{m}}{2}\ket{0}\left[ \theta(z-(b-\frac{a}{2}))-\theta(z-(b+\frac{a}{2}))\right] A_{an}~ \sin\big(k_{an}(b+\frac{a}{2}-z)\big)\nonumber\\
-\cos\frac{\theta_{m}}{2}e^{i\phi_{m}}\ket{1}\left[ \theta(z+(b+\frac{a}{2}))-\theta(z+(b-\frac{a}{2}))\right] A_{an}~\sin\big(k_{an}(b+\frac{a}{2}+z)\big)
\label{unperturbed state inside finite well}
\end{eqnarray}
\end{widetext}
In the limit $V_0\rightarrow\infty$, we obtain $\ket{\phi_{sn}(z)}\rightarrow\ket{\Phi_{0sn}(z)}$ and $\ket{\phi_{an}(z)}\rightarrow\ket{\Phi_{0an}(z)}$ as required. One can verify that $\inpr{\phi_{ak}(z)}{\phi_{sn}(z)}=0$, for all $k,n$. However, $\inpr{\phi_{sk}(z)}{\phi_{sn}(z)}\ne 0$ and $\inpr{\phi_{ak}(z)}{\phi_{an}(z)}\ne 0$, for $k\ne n$. We can make them zero via Grahm-Schmidt orthogonalization. Arbitrary constants $A_{sn},A_{an},B_{sn}$ and $B_{an}$ can be fixed by the requirement that each of the states $\left\lbrace \ket{\phi_{sn}(z)},\ket{\phi_{ak}(z)}\right\rbrace$, for all $s,k$, be normalised and satisfy completeness relation along with the states $\left\lbrace \ket{\phi'_{l}(z)}\right\rbrace $, where $\ket{\phi'_{l}(z)}=\ket{\hat{m}}\sqrt{\frac{2}{2b+a}}\cos(\frac{l\pi}{2b+a}z)$ is an eigenstate of $H_{S}$ with eigenvalue $E'_{l}>V_{0}$. Also $\inpr{\phi'_{l}(z)}{\phi_{sn,an}(z)}\ne 0$. Taking the state of our main interest, $\ket{\phi_{s1}(z)}$, as reference, we can orthogonalize the complete set $\left\lbrace \ket{\phi_{sk}(z)},\ket{\phi_{ak}(z)},\ket{\phi'_{l}(z)}\right\rbrace$ via Grahm-Schmidt orthogonalization. The set $\left\lbrace \ket{\phi_{sk}(z)},\ket{\phi_{ak}(z)},\ket{\phi'_{l}(z)}\right\rbrace$, after orthonormalization, should satisfy the completeness relation analogous to that in eq. \ref{compltness rel}, i.e., 
\begin{eqnarray}
\int\limits_{-\infty}^{\infty}dz'\bigg[\sum\limits_{k=1}^{M}\bigg(\outpr{\phi_{sk}(z)}{\phi_{sk}(z')}+\outpr{\phi_{ak}(z)}{\phi_{ak}(z')}\bigg)+~~~~~\nonumber\\\sum\limits_{l=2M+1}^{\infty}\outpr{\phi'_{l}(z)}{\phi'_{l}(z')}\bigg]=\int\limits_{-\infty}^{\infty}dz'\sum\limits_{i=1}^{\infty}\outpr{\phi''_{i}(z)}{\phi''_{i}(z')}=\mathbbm{1}~~~~~~
\label{completnes of phi_sn}
\end{eqnarray}
\begin{eqnarray}
\mathrm{where},~\ket{\phi_{sk}(z)}=\ket{\phi''_{2k-1}(z)},~\ket{\phi_{ak}(z)}=\ket{\phi''_{2k}(z)},\nonumber\\\ket{\phi'_{l}(z)}=\ket{\phi''_{l}(z)}~~~~~~~~~~~~~~~
\label{ket phi'' defined}
\end{eqnarray}
We can justify this by the fact that, in the limit $V_{0}\rightarrow\infty$ we recover the completeness relation in eq. \ref{compltness rel}, which corresponds to $V_{I}(z)$ (\ref{box potential defined}). By perturbing $V_{I}$ with a small negative potential, we obtain the case of finite potential, $V_{F}(z)$ (\ref{H_S defined}). As perturbation doesnot destroy the completeness property of the set of states being perturbed, completeness relation in eq. \ref{completnes of phi_sn} is justified. 

\textbf{Protective Measurement:} An important requirement to do protective measurement is, system should be initially in a nondegenerate eigenstate of system Hamiltonian, $H_{S}$. Energy of system electron, just before trapping it in double well potential is:
\begin{eqnarray}
\frac{1}{2M}(p_{0z}\pm\Delta p_{z})^2=\frac{1}{2M}(\gamma B_{i}\frac{\hbar}{2}\tau\pm\Delta p_{z})^2\nonumber\\=\frac{1}{2M}(\gamma B_{i}\frac{\hbar}{2}\tau)^2+\Delta E
\label{enrgy cr.ing variance}
\end{eqnarray}
where, $p_{0z}$ is the average momentum, $\Delta p_{z}$ is the standard deviation in momentum and 
\begin{equation}
\Delta E=\frac{1}{2M}(\Delta p_{z}^{2}\pm 2\gamma B_{i}\frac{\hbar}{2}\tau\Delta p_{z})
\label{Delta E}
\end{equation}
is the uncertainty in kinetic energy of system electron. Let us consider the ground state $\ket{\phi_{s1}(z)}$ (\ref{state inside finite well}) (as $E_{sn}$ is less than $E_{an}$ \cite{QM_Cohen}). From eq. \ref{E reltd to k}, $E_{s1}=\frac{\hbar^{2}k^{2}_{s1}}{2M}$. But eq. \ref{constraint eq sym} says, $k_{s1}$ depends on the potential well parameters $a,b,V_{0}$. To push the system electron into the state $\ket{\phi_{s1}(z)}$, following condition must be satisfied,
\begin{eqnarray}
E_{s1}(a,b,V_0)=\frac{1}{2M}(\gamma B_{i}\frac{\hbar}{2}\tau)^2
\label{E_sn=KE of sys e-}
\end{eqnarray}
Even if the values of $a,b,V_0$ in $E_{s1}(a,b,V_0)$ are fixed by some requirements like: compensating increase in $\Delta z$ as $\Delta p_z$ is squeezed (see below eq. \ref{ket phi_dwp}), etc., constraint eq. \ref{E_sn=KE of sys e-} can still be satisfied by varying the free parameters $B_i$ and $\tau$. Hence, we can say, energy of system electron, just before trapping is $E_{s1}+\Delta E$. In general, state of system electron after trapping it in double well potential will be,
\begin{eqnarray}
\ket{\phi_{dwp}}=c_{1}\ket{\phi_{s1}(z)}+\sum\limits_{i\neq 1}c_{i}\ket{\phi''_{i}(z)}
\label{ket phi_dwp}
\end{eqnarray}
where, $\ket{\phi''_{i}(z)}$ is given by eq. \ref{ket phi'' defined}. However, system electron starts with zero average momentum as mentioned below eq. \ref{m cap decomposed}. At $t=0$, $\Delta z\Delta p_z=\hbar/2$. In principle we can start at $t=0$ with squeezed wave packet such that $\Delta p_z\rightarrow 0$. Switch on the inhomogeneous magnetic field for an interval of time $\tau$ such that it gains an average kinetic energy $E_{s1}(a,b=a/2+\epsilon,V_0)$ ($\epsilon>0$, to avoid blowing up of cot hyperbolic function in eq. \ref{constraint eq sym}) as given by the constraint eq. \ref{E_sn=KE of sys e-}. Now adiabatically change the Hamiltonian from $p_z^2/(2M)$ to $H_S$ (\ref{H_S defined}) with $b=a/2+\epsilon$. By this, energy of system electron will not change, as we are introducing potential barriers only. Width `$a$' of well is suitably chosen so as to take care of increase in $\Delta z$ due to squeezing of $\Delta p_z$. In principle `$a$' can be arbitrarily large. With this we obtain: $\Delta E\rightarrow 0$, $|c_{1}|\rightarrow 1$ and $|c_{i}|\rightarrow 0$ ($i\neq 1$) with state of system electron being $\ket{\phi_{s1}(z)}$ corresponding to $b=a/2+\epsilon$ \footnote[2]{This can also be justified as follows: Consider a Box normalized plane wave: $u_n(z)=L^{-1/2}\exp(ik_{n_z}z)$ with periodic boundary condtion. It is the spatial wave function of system electron. It is an eigenstate of $p_z^2/(2M)$ with discrete energy $E^f_{n_z}=\hbar^2 k_{n_z}^2/(2M)$, where, $k_{n_z}=2\pi n_z/L$, $n_z=0,\pm 1,\pm 2,..$. $L$ is the length of box, which is large but finite. Spacing between the energy levels can be made as small as we want by increasing $L$ \cite{QM_schiff}. Now switch on the inhomogeneous magnetic field for an interval of time $\tau$, such that $\frac{1}{2M}(\gamma B_{i}\frac{\hbar}{2}\tau)^2=E^f_{m_z}=E_{s1}(a,b=a/2+\epsilon,V_0)$. Change the Hamiltonian of system electron from $p_z^2/(2M)$ to $H_S$ with $b=a/2+\epsilon$, in an interval of time $T^f>>\hbar/|E^f_{m_z}-E^f_{m_z+1}|$. Then, by adiabatic theorem (see below eq. \ref{eigenvale E tilde def}), system electron will end up in the eigenstate of $H_S$: $|\phi_{s1}(z)>$ corresponding to $b=a/2+\epsilon$.}. Now again adiabatically change the Hamiltonian from $H_S$ with $b=a/2+\epsilon$ to $H_S$ with $b=a/2+\epsilon+\mathcal{M}$, where $\mathcal{M}(>0)$ is large but finite i.e., separate the two wells very slowly. Hence, in principle, `$b$' can also be made arbitrarily large. Here, energy of system electron changes. Hence, by adiabatic principle (see below eq. \ref{eigenvale E tilde def}), system electron will end up in the state $\ket{\phi_{s1}(z)}$ corresponding to $b=a/2+\epsilon+\mathcal{M}$. Hence, \textit{in principle, we can push the system electron into the nondegenerate eigenstate,} $\ket{\phi_{s1}(z)}$, \textit{of system Hamiltonian} $H_{S}$. 

Or, it is sufficient for the system electron to have average kinetic energy (which is $\frac{1}{2M}(\gamma B_{i}\frac{\hbar}{2}\tau)^2$) sufficiently greater than the ground state energy $E_{s1}$, which is exactly known. As, all systems have natural tendency to settle down to their ground state, as it is most stable and has least possible energy, if we wait sufficiently long, system electron will settle down to the ground state $\ket{\phi_{s1}(z)}$ \cite{prot_meas_critq_haridas}\footnote[10]{To settle down to ground state, system electron has to emit photon(s). But system electron is already entangled with it's spin d.o.f. Hence emission of photon may cause mutual entanglement between spin and spatial d.o.f of system electron, and photon. Then it may be necessary to measure protectively the combined three systems, which we are not considering here}. Hence, in this method we need not take the limit $\Delta p_{z}\rightarrow 0$, unlike in the previous method. Hence, by either of these two methods, we can push the system electron into a nondegenerate eigenstate $\ket{\phi_{s1}(z)}$ (\ref{state inside finite well}) of system Hamiltonian $H_{S}$ (\ref{H_S defined}). Even if we trap system electron in the eigenstate $\ket{\phi_{sn}(z)}$, $n>1$, by satisfying the constriant: $E_{sn}=\frac{1}{2M}(\gamma B_{i}\frac{\hbar}{2}\tau)^2$, it won't be stable due to vaccum fluctuations and ultimately decays to ground state, $\ket{\phi_{s1}(z)}$.

We now carry out  protective measurement on system electron trapped in the double well potential,  using a probe electron which passes in-between the two wells. As it passes, it interacts with system electron through coulomb potential energy. Even though we have trapped system electron in 1-D potential well, it can still interact with probe electron moving in $x$-$z$ plane(let positive $x$-axis be along positive $V_{I}$-axis in Fig. \ref{fig_double_well}), via coulomb potential energy. Trapped electron is analogous to line charge density(Coulomb per meter), where the potential well is opaque to matter wave(hence charges cannot escape from metal surface) but transparent to electric field and hence they can interact with other free charges outside the well. Only trapped system electron sees infinite potential barrier in between the two wells, but not the probe electron \footnote[13]{For eg., imagine a three dimensional (3D) double box potential, where, potential inside the box is zero, while infinite on the six walls of box. Box is opaque to matter wave (as trapped electron cannot escape), while transparent to electromagnetic field. Now, to break the degeneracy, connect the two boxes with a long tube, such that it does not obstruct the motion of probe electron. Inside the tube, potential is finite ($V_0$), whereas infinite on its walls. This is analogous to the potential $V_F(z)$ (eq. \ref{H_S defined}). Our protocol can be generalized to 3D double box potential case.}.\\ As stated in \cite{prot_measure}, adiabatic condition required to do protective measurement will be satisfied, assuming system electron is initially in the nondegenerate energy eigenstate $\ket{\phi_{sn}(z)}$, provided the probe electron crosses the potential well in a time $T$, which is large compared to $\hbar /\left| E_{sn}-E_{an}\right|$ where, $E_{sn}$, $E_{an}$ are the energies of the states $\ket{\phi_{sn}(z)}$, $\ket{\phi_{an}(z)}$ respectively. As the probe electron is moving very slowly, we can neglect the magnetic field induced due to motion of charge. In a more rigorous calculation we can treat it as low energy Quantum Electrodynamical (QED) problem. However, here we are going to approximate it as quantum electrostatic problem. Let the probe electron have an initial constant momentum, $p_{a0x}\hat{i}$, where $\hat{i}$ is an unit vector along $x$-axis. As it measures the system electron protectively, it gains component of momentum along $z$-axis, and hence drifts along, say, positive $z$-axis. As a result, potential energy of interaction, $H^a_{int}=\frac{1}{4\pi\epsilon_{0}}\frac{e^{2}}{\left| \vec{r}-\vec{r}_{a}(t)\right|}$, changes, where $\vec{r}$ is the position vector of system electron and $ \vec{r}_{a}(t)$ that of probe electron. However, if the separation between wells, $2(b-a/2)$, is sufficiently large (as justified below eq. \ref{ket phi_dwp}) and time interval $T$ relatively small (i.e., $T>>\hbar /\left| E_{s1}-E_{a1}\right|$ still holds) \footnote[14]{As a first step, to get an approximate solution, we are assuming that $T$ is relatively small, so that $H^a_{int}$ becomes time independent. In a more accurate solution we can treat $H^a_{int}$ as time dependent perturbation or handle it as a quantum electrodynamical problem, and we can handle arbitrarily large $T$.}, we can neglect the change in $H^a_{int}$ during the time interval $T$ and approximate it as: 
\begin{equation}
H^a_{int}\approxeq\frac{1}{4\pi\epsilon_{0}}\frac{e^{2}}{\left|z-z_{a}\right|}
\label{H_int approx}
\end{equation}
with, center of wave packet of probe electron: $z_{a0}=0$. It is as if probe electron is fixed exactly in between the two potential wells. \\As the separation between two wells, $2(b-a/2)$, can be made arbitrarily large as justified below eq. \ref{ket phi_dwp}, $1/\left|z-z_{a}\right|$ can be made arbitrarily small for every $z$ such that: $(b-a/2)\le z\le (b+a/2)$ and $-(b+a/2)\le z\le -(b-a/2)$. Hence $H^a_{int}$ can be made arbitrarily small compared to $\left| E_{sn}-E_{an}\right|$ (also see last para of \footnote[9]{Initial state of probe electron (eq. \ref{ket zeta(T) ini}) was a Gaussian minimum (i.e., $\Delta z_a\Delta p_{az}=\hbar/2$) wave packet centered at $p_{a0z}=0$. Because $p_{a0z}=0$, at any time $t$: $z_{a0}=0$ (see below eq. \ref{genrtr of momentum}). We did calculations from eq. \ref{final momentm gained}-\ref{p_aoz,final} assuming the spread $\Delta z_a$ to be small. However, in principle $\Delta z_a$ can be arbitrarily large, as the separation between potential wells can also be made arbitrarily large (below eq. \ref{ket phi_dwp}) and one can handle it as QED problem or calculate the integral (\ref{genrtr of momentum}) to all orders in $\epsilon$ (spreading). Hence, in principle we can start with a squeezed minimum wave packet: $\Delta p_{az}\rightarrow 0$ with $\Delta z_a\Delta p_{az}=\hbar/2$, $p_{a0z}=0$ and $z_{a0}=0$. After the interaction with system electron, to measure the momentum gained by probe electron, shine it with a photon of wavelength $\lambda$. By measuring the change in frequency, due to Doppler effect, of scattered photon we can calculate the velocity of probe electron. The change in momentum of probe electron ($=\Delta p'_{az}$, which is also the error in momentum measurement) due to collision with photon, is inversely proportional to $\lambda$ (Compton effect) \cite{quant_theory_mesurmnt_Wheeler_zurek}. In principle it is possible to take: $\lambda\rightarrow\infty$. As a result, error: $\Delta p'_{az}\rightarrow 0$ and the measured value of momentum of probe electron tends to $p_{a0z}^{final}$. Of course, error in position measurement: $\Delta z'_a\rightarrow\infty$ (as $\Delta z'_a\Delta p'_{az}\ge\hbar/2$), in which any way we are not interested. 

Also, even in the limit $\Delta p_{az}\rightarrow 0$, it is possible to make $H^a_{int}$ (\ref{H_int approx}) goto zero as follows: Let the wave packet of probe electron be centered at $z_{a0}=0$ with standard deviation $\Delta z_a$. Then, maximum possible strength of $H^a_{int}$ is proportional to $1/(b-a/2-\Delta z_a)$. As the separation between two potential wells can be made arbitrarily large, we can take $b-a/2=n\Delta z_a,~n>1$. Now, in the limit $\Delta z_a\rightarrow\infty$ (i.e., $\Delta p_{az}\rightarrow 0$), $H^a_{int}\rightarrow 0$. We also note that, the requirement $\Delta p_z\rightarrow 0$ (hence, $\Delta z\rightarrow\infty$) (below \ref{ket phi_dwp}) demands width `$a$' to be of the order of $\Delta z$. Hence, $b\approx \Delta z/2+n\Delta z_a$. In the limit $\Delta z\rightarrow\infty$ and $\Delta z_a\rightarrow\infty$, $b$ also tends to infinity. This can be satisfied as $b$ can be arbitrarily large.}, and second para of \footnote[6]{Let $\lambda$ be the strength of $H^a_{int}$ (relation \ref{H_int approx}). In the nondegenerate perturbation theory, first order correction to $|\phi_{s1}>$ goes as $\lambda/(E_{s1}-E_{a1})$ (p1101 of \cite{QM_Cohen}). In the limit $|E_{s1}-E_{a1}|\rightarrow 0$ (i.e., $V_0\rightarrow \infty$) evanescent wave goes to zero. As explained below relation \ref{H_int approx}, $\lambda$ depends on separation between two wells, which can be made arbitrarily large, and hence $\lambda$ arbitrarily small. We can choose the separation between two wells such that $\lambda$ is numerically proportional to $|E_{s1}-E_{a1}|^n,~n>1$. Now, in the limit $|E_{s1}-E_{a1}|\rightarrow 0$, both  first order correction to $|\phi_{s1}>$ and evanescent wave goes to zero. Similarly we can prove all higher order corrections to $|\phi_{s1}>$ (p154 of \cite{QM_schiff}) also goes to zero. 

Also, criterion to treat $H^a_{int}$ as a small perturbation to $H_S$ is (p1096 \cite{QM_Cohen}): $\lambda<<|E_{s1}-E_{a1}|$. This is also satisfied by choosing $\lambda$ numerically proportional to $|E_{s1}-E_{a1}|^n,~n>1$ and taking the limit $|E_{s1}-E_{a1}|\rightarrow 0$. 

Also, in the footnote below eq. \ref{p_aoz,final}, we took numerical value of $\lambda=1/(b-a/2-\Delta z_a)$ (upper limit of $\lambda$). Requirement that $H^a_{int}$ should goto zero, gave: $b-a/2=n\Delta z_a,~n>1$. Equating this with above numerical value of $\lambda=|E_{s1}-E_{a1}|^m,~m>1$, we obtain: $\Delta z_a|E_{s1}-E_{a1}|^m=1/(n-1)$. This puts a constraint on the rate at which $\Delta z_a$ should goto infinity or $|E_{s1}-E_{a1}|$ should goto zero i.e., the rates should be such that the product $\Delta z_a|E_{s1}-E_{a1}|^m$ remains finite ($=1/(n-1)$). This can always be satisfied, for eg., taking numerical value of  $|E_{s1}-E_{a1}|^m=1/((n-1)\Delta z_a)$. 

Similarly, adiabaticity condition: $T>>\hbar/|E_{s1}-E_{a1}|$ (below eq. \ref{eigenvale E tilde def}), can also be satisfied even in the limit $|E_{s1}-E_{a1}|\rightarrow 0$, by choosing $T$, which can also be arbitrarily large, numerically proportional to $|E_{s1}-E_{a1}|^{-n},~n>1$. Hence, evanescent wave is not a problem while recombining.}), there by satisfying the weak interaction requirement for protective measurement, i.e., $H^a_{int}$ is so weak that it cannot cause transition between the states $\ket{\phi_{sn}}$ and $\ket{\phi_{an}}$. We can vary $\left| E_{sn}-E_{an}\right|$ by varying $V_{0}$. 

As we are interested only in the change in $z$-component of probe electron's momentum, we can neglect its other components. With this and the electroststic approximation in eq. \ref{H_int approx}, we can take free evolution Hamiltonian of probe electron, $H_{a}=p_{az}^2/(2M)=0$ \footnote[16]{Even if we start with probe electron being a Gaussian minimum (i.e., $\Delta z_a\Delta p_{az}=\hbar/2$) wave packet, with $p_{a0z}=0$, $z_{a0}=0$ and $\Delta p_{az}\rightarrow 0$ (see footnote below eq. \ref{p_aoz,final}), as it interacts with system electron, it gains momentum and hence $H_a$ is no more zero. However, if $H_S+H_{int}>>H_a$ (which might be possible by suitably choosing the free parameters $a,~b$ and $V_0$ in $E_{s1}(a,b,V_0)$ (\ref{E_sn=KE of sys e-}). However, we note that $a$ is constrained by the requirement $\Delta p_z\rightarrow 0$, $b$ by the weak measurement criterion and $V_0$ by adiabaticity requirement. As $a,~b$ and $V_0$ can be arbitrarily large, inspite of these constraints there might be a possibility of choosing them such that $H_S+H_{int}>>H_a$ is also satisfied.), then we can neglect evolution under $H_a$. In an exact calculation one can take into account $H_a$ also (eg., treating it as a QED problem or following a procedure described in \cite{prot_meas_critq_haridas}) but the conclusions remain unaffected, because, presence of $H_a$ cannot cause entanglement between system and probe electrons. $H_a$ may cause wave packet to spread, which can be taken into account.}. Hence, System and probe electrons evolve under the Hamiltonian:
\begin{eqnarray}
H=H_{S}+\frac{1}{4\pi\epsilon_{0}}\dfrac{e^2}{|z-z_{a}|}
\label{H defined}
\end{eqnarray}
where, $H_{S}$ is the system Hamiltonian given in eq. \ref{H_S defined}, $z$ is the position coordinate corresponding to system electron and $z_{a}$ that of probe electron. State of combined system at time $t=T$ is:
\begin{eqnarray}
\ket{\zeta(T)}=\exp\bigg(\frac{-i}{\hbar}(H_{S}+\frac{1}{4\pi\epsilon_{0}}\dfrac{e^2}{|z-z_{a}|})T\bigg)\ket{\phi_{s1}(z)}\bar{\Theta}(p_{az})\nonumber\\
\label{ket zeta(T) ini}
\end{eqnarray}
where, $\ket{\phi_{s1}(z)}$ is given by eq. \ref{state inside finite well}, $\bar{\Theta}(p_{az}-0)$ is the Gaussian wave packet of probe electron in momentum space centered at $p_{a0z}=0$. Following Hari Dass and Tabish \cite{prot_meas_critq_haridas}, we insert completeness relation similar to that in eq. \ref{completnes of phi_sn}, to obtain:
\begin{widetext}
\begin{eqnarray}
\ket{\zeta(T)}=\int\limits_{-\infty}^{\infty}dz_{a}\int\limits_{-\infty}^{\infty}dz'\sum\limits_{k=1}^{\infty}\exp\bigg(\frac{-i}{\hbar}\bigg(H_{S}+\frac{1}{4\pi\epsilon_{0}}\dfrac{e^2}{|\hat{z}-\hat{z}_{a}|}\bigg)T\bigg)\delta(z_{a})(\outpr{\tilde{\phi}_{k}(z)}{\tilde{\phi}_{k}(z')})\ket{\phi_{s1}(z')}\bar{\Theta}(p_{az})
\label{ket zeta comp insertd}
\end{eqnarray}
\end{widetext}
where, hat on $z,z_{a}$ stands for operator, $\delta(z_a)$ is the Dirac delta function and $\ket{\phi_{s1}(z)}$ has been changed to $\ket{\phi_{s1}(z')}$ because see\footnote[7]{An arbitrary function inside the single potential well can be expanded as:$f(z)=\int\limits_{0}^{a}dz'f(z')\delta(z'-z)$. Substituting for $\delta(z'-z)$ from eq.\ref{delta fn} and rearranging, we get the required expansion. Note that $z$ needs to be changed to $z'$.}. If, $\hat{A}\ket{a_{i}}=a_{i}\ket{a_{i}}$ then $f(\hat{A})\ket{a_{i}}=f(a_{i})\ket{a_{i}}$. Using this property we can push $\delta(z_{a})$ to the extreme left. Then we obtain:
\begin{eqnarray}
\bigg(H_{S}+\frac{1}{4\pi\epsilon_{0}}\dfrac{e^2}{|\hat{z}-z_{a}|}\bigg)\ket{\tilde{\phi}_{k}(z)}=\tilde{E}_{k}(z_{a})\ket{\tilde{\phi}_{k}(z)},\mathrm{where},\nonumber\\\tilde{E}_{k}(z_{a})=\expec{H_{S}}_{\tilde{\phi}_{k}}+\frac{1}{4\pi\epsilon_{0}}e^{2}\bra{\tilde{\phi}_{k}(z)}\dfrac{1}{\left|z-z_{a}\right|}\ket{\tilde{\phi}_{k}(z)}~~~~~~~
\label{eigenvale E tilde def}
\end{eqnarray}
As discussed in the beginning of `Protective Measurement', criteria of adiabaticity and weak interaction between system and probe electrons are satisfied. According to Adiabatic theorem, at $t=0$ if the system electron is in a nondegenerate eigenstate of the system Hamiltonian $H_S$ (which we take to be the ground state, $\ket{\phi_{s1}(z)}$), then, if $H_S$ changes very slowly to $H=H_{S}+H^a_{int}$ at $t=T$, such that $T>>(\hbar/\left| E_{s1}-E_{a1}\right|)$ (adiabatic approximation, eq. 10.15 of \cite{QM_Griffiths}), then $\ket{\phi_{s1}(z)}$ changes to the corresponding nondegenerate eigenstate of $H$ (with $\hat{z}_a$ replaced by $z_a$, as in eq. \ref{eigenvale E tilde def}) i.e., ground state $\ket{\tilde{\phi}_{1}(z)}$ (pp327-29 of \cite{QM_Griffiths}). This is true for arbitrary $H^a_{int}$ (see the generalization on p330 of \cite{QM_Griffiths}). Further, as explained below eq. \ref{H_int approx}, if the separation between wells is arbitrarily large, then, $H^a_{int}$ will be arbitrarily small (also see \footnote[6]{}). Then, by nondegenerate perturbation theory, correction to all orders to the state $\ket{\phi_{s1}(z)}$, due to perturbation $H^a_{int}$, goes to zero \cite{QM_schiff} and we obtain: $\ket{\tilde{\phi}_{1}(z)}$ tends to $\ket{\phi_{s1}(z)}$. Generalizing to an arbitrary energy eigenstate, we obtain: $\ket{\tilde{\phi}_{k}(z)}$ tends to $\ket{\phi''_{k}(z)}$, where $\ket{\phi''_{k}(z)}$ is defined in eq. \ref{ket phi'' defined}. Using orthonormality of states, i.e., 
\begin{eqnarray}
\int\limits_{-\infty}^{\infty}dz'\inpr{\phi''_{i}(z')}{\phi_{sk}(z')}=
\begin{cases}
    1,~\mathrm{if}~i=2k-1,k\le M\\
    0,~\mathrm{otherwise}\\
\end{cases}\nonumber\\
\int\limits_{-\infty}^{\infty}dz'\inpr{\phi''_{i}(z')}{\phi_{ak}(z')}=
\begin{cases}
    1,~\mathrm{if}~i=2k,k\le M\\
    0,~\mathrm{otherwise}\\
\end{cases}\nonumber\\
\int\limits_{-\infty}^{\infty}dz'\inpr{\phi''_{i}(z')}{\phi'_{l}(z')}=
\begin{cases}
    1,~\mathrm{if}~i=l,l\ge 2M+1\\
    0,~\mathrm{otherwise}\\
\end{cases}
\label{orthonormlty of phi''} 
\end{eqnarray}
and using eq.\ref{eigenvale E tilde def}, $\ket{\zeta(T)}$ in eq. \ref{ket zeta comp insertd} reduces to,
\begin{widetext}
\begin{eqnarray}
\ket{\zeta(T)}=\int\limits_{-\infty}^{\infty}dz_{a}\delta(z_{a})\exp\big(\frac{-i}{\hbar}\expec{H_{S}}_{\phi_{s1}}T\big)\exp\bigg(\frac{-i}{\hbar}\frac{1}{4\pi\epsilon_{0}}e^{2}\bra{\phi_{s1}(z)}\dfrac{1}{\left| z-z_{a}\right|}\ket{\phi_{s1}(z)}T\bigg)\ket{\phi_{s1}(z)}\bar{\Theta}(p_{az})\nonumber\\=\exp\big(\frac{-i}{\hbar}\expec{H_{S}}_{\phi_{s1}}T\big)~\ket{\phi_{s1}(z)}\exp\bigg(\frac{-i}{\hbar}\frac{1}{4\pi\epsilon_{0}}e^{2}\bra{\phi_{s1}(z)}\dfrac{1}{\left| z-\hat{z} _{a}\right|}\ket{\phi_{s1}(z)}T\bigg)\bar{\Theta}(p_{az})
\label{ket zeta final}
\end{eqnarray}
\end{widetext}
\begin{equation}
\mathrm{Let},~f(\hat{z}_{a})=\bra{\phi_{s1}(z)}\dfrac{1}{\left| z-\hat{z} _{a}\right|}\ket{\phi_{s1}(z)}
\label{f(z_a defined)}
\end{equation}
Using the definition of wavepacket (p1462 of \cite{QM_Cohen}), we obtain:
\begin{eqnarray}
\exp\bigg(\frac{-i}{\hbar}\frac{1}{4\pi\epsilon_{0}}e^{2}f(\hat{z}_{a})T\bigg)\bar{\Theta}(p_{az})=~~~~~~~~~~~~\nonumber\\\frac{1}{\sqrt{2\pi\hbar}}\int\limits_{-\infty}^{\infty}dz_{a}\Theta(z_{a})\exp\bigg(\frac{-i}{\hbar}\bigg(\frac{1}{4\pi\epsilon_{0}}e^{2}f(z_{a})T+p_{az}z_{a}\bigg)\bigg)~~~~~~
\label{genrtr of momentum}
\end{eqnarray}
Wave packet of probe electron in position space is centered at $z_{a0}=0$ (see below eq. \ref{H_int approx}) and center won't change with time, as $p_{a0z}=0$ (see above eq. \ref{splitting stage}). Expanding $f(z_{a})$ around $z_{a}=z_{a0}=0$, we obtain, $f(z_{a})=f(0)+f'(0)z_{a}+O(z^{2}_{a})$, where, $f'(0)=\frac{d}{dz_{a}}f(z_{a})\mid_{z_{a}=0}$. If the spreading of wave packet is small during the time interval $T$ \footnote[12]{However we note that, in principle it is possible to take into account, complete spreading of wave packet during the time interval $T$. For eg., if we treat it as a QED problem , we get exact solution or do the integral (\ref{genrtr of momentum}) to all orders in $\epsilon$. However, to get an approximate solution, we consider spreading of wave packet only to first order in $\epsilon$.}\cite{prot_measure}, $\Theta(z_{a})$, a Gaussian wave packet, is appreciable only in the interval: $-\epsilon <z_{a}<\epsilon$, $\epsilon>0$, such that $\epsilon^{2}$ and higher order terms are negligible i.e., $\Theta(\epsilon+\delta)\approxeq 0$, where, $\delta>0$. Hence in integral (\ref{genrtr of momentum}), $f(z_{a})$ can be approximated as: $f(z_{a})\approxeq f(0)+f'(0)z_{a}$. Eq. \ref{genrtr of momentum} reduces to \footnote[3]{$\int\limits_{-\infty}^{\infty}dz_{a}\Theta(z_{a})\exp\bigg(\frac{-i}{\hbar}\bigg(\frac{1}{4\pi\epsilon_{0}}e^{2}f(z_{a})T+p_{az}z_{a}\bigg)\bigg)\approxeq \int\limits_{-\epsilon}^{\epsilon}dz_{a}\Theta(z_{a})\exp\bigg(\frac{-i}{\hbar}\bigg(\frac{1}{4\pi\epsilon_{0}}e^{2}f(z_{a})T+p_{az}z_{a}\bigg)\bigg)\approxeq \int\limits_{-\epsilon}^{\epsilon}dz_{a}\Theta(z_{a})\exp\bigg(\frac{-i}{\hbar}\bigg(\frac{1}{4\pi\epsilon_{0}}e^{2}(f(0)+f'(0)z_{a})T+p_{az}z_{a}\bigg)\bigg)\approxeq \int\limits_{-\infty}^{\infty}dz_{a}\Theta(z_{a})\exp\bigg(\frac{-i}{\hbar}\bigg(\frac{1}{4\pi\epsilon_{0}}e^{2}(f(0)+f'(0)z_{a})T+p_{az}z_{a}\bigg)\bigg)$}:
\begin{eqnarray}
\exp\bigg(\frac{-i}{\hbar}\frac{1}{4\pi\epsilon_{0}}e^{2}f(\hat{z}_{a})T\bigg)\bar{\Theta}(p_{az})=~~~~~~~~~~~~\nonumber\\\exp\bigg(\frac{-i}{\hbar}\frac{1}{4\pi\epsilon_{0}}e^{2}f(0)T\bigg)\bar{\Theta}\bigg(p_{az}+\frac{1}{4\pi\epsilon_{0}}e^{2}f'(0)T\bigg)
\label{final momentm gained}
\end{eqnarray}
Finally we obtain,
\begin{eqnarray}
\ket{\zeta(T)}=\exp\big(\frac{-i}{\hbar}\bigg(\expec{H_{S}}_{\phi_{s1}}+\frac{1}{4\pi\epsilon_{0}}e^{2}f(0)\bigg)T\big)\nonumber\\\ket{\phi_{s1}(z)}\bar{\Theta}\bigg(p_{az}+\frac{1}{4\pi\epsilon_{0}}e^{2}f'(0)T\bigg)
\label{final most state of sys+probe}
\end{eqnarray}
State of system electron is unaltered while probe electron has gained an average momentum: $-\frac{1}{4\pi\epsilon_{0}}e^{2}f'(0)T$, and we are now going to show that $f'(0)$ is proportional to $\cos\theta_{m}$. Substituting eq.\ref{state inside finite well} into eq.\ref{f(z_a defined)}, we obtain: 
\begin{eqnarray}
f(z_{a})=\bra{\phi_{0s1}(z)}\dfrac{1}{\left| z-z _{a}\right|}\ket{\phi_{0s1}(z)}\nonumber\\+\int\limits_{-(b-\frac{a}{2})}^{b-\frac{a}{2}}dz~B^{2}_{s1}\dfrac{\cosh^{2}(q_{s1}z)}{\left| z-z _{a}\right|}=h(z_{a})+g(z_{a})
\label{f=h+g}
\end{eqnarray}
In eq.s \ref{f=h+g} and \ref{h(z_a) evaluated}, integration variable is $z$, and hence $z_a$ is a constant as far as integration is concerned. We want to know the rate at which the function $f(z_{a})$ changes, w.r.t $z_{a}$, at $z_{a}=0$. Hence w.r.t $z_a=0$ (the value of our interest) we can write: 
\begin{eqnarray}
\left| z-z _{a}\right|=
\begin{cases}
    z-z _{a},~\mathrm{for}~z\ge 0\\
    -(z-z _{a}),~\mathrm{for}~z<0\\
\end{cases}
\label{abs z-za defined} 
\end{eqnarray}
However, we want $f'(0)$ but not $f(0)$. Hence, we are going to first differentiate $f(z_a)$ w.r.t $z_a$ and then put $z_a=0$. Using (\ref{abs z-za defined}) and substituting for $\ket{\phi_{0s1}(z)}$ from eq.\ref{unperturbed state inside finite well} and integrating w.r.t $z$ from $-\infty$ to $+\infty$, we obtain,
\begin{eqnarray}
h(z_{a})=\bra{\phi_{0s1}(z)}\dfrac{1}{\left| z-z _{a}\right|}\ket{\phi_{0s1}(z)}\nonumber\\=A_{s1}^{2}\cos^{2}\frac{\theta_{m}}{2}\int\limits_{b-\frac{a}{2}}^{b+\frac{a}{2}}dz~\dfrac{\sin^{2}(k_{s1}(b+\frac{a}{2}-z))}{z-z_{a}}\nonumber\\+A_{s1}^{2}\sin^{2}\frac{\theta_{m}}{2}\int\limits_{b-\frac{a}{2}}^{b+\frac{a}{2}}dz~\dfrac{\sin^{2}(k_{s1}(b+\frac{a}{2}-z))}{z+z_{a}}
\label{h(z_a) evaluated}
\end{eqnarray}
Differentiating $h(z_{a})$ w.r.t $z_{a}$ in eq. \ref{h(z_a) evaluated}, using Leibnitz rule for differentiation under the integral sign, and then taking $z_{a}=0$, we obtain,
\begin{eqnarray}
h'(0)=\cos\theta_{m}~A_{s1}^{2}\int\limits_{b-\frac{a}{2}}^{b+\frac{a}{2}}dz~\dfrac{\sin^{2}(k_{s1}(b+\frac{a}{2}-z))}{z^{2}}~~~~
\label{h'(0) evaluated}
\end{eqnarray}
Similarly,
\begin{eqnarray}
g(z_{a})=B^{2}_{s1}\lim\limits_{\epsilon\rightarrow 0}\int\limits_{\epsilon}^{b-\frac{a}{2}}dz\bigg(\dfrac{\cosh^{2}(q_{s1}z)}{z+z_{a}}+\dfrac{\cosh^{2}(q_{s1}z)}{z-z_{a}}\bigg)~~~~~
\label{g(z_a)evaluated}
\end{eqnarray}
To avoid pole at $z=z_{a}=0$, we have introduced the limit. As before, differentiating $g(z_{a})$ in eq. \ref{g(z_a)evaluated}, w.r.t $z_{a}$, and then taking $z_{a}=0$, we get $g'(0)=0$. $\therefore$ using eq. \ref{f=h+g}, we obtain $f'(0)=h'(0)$ (\ref{h'(0) evaluated}). From eq. \ref{final most state of sys+probe} we obtain the final average momentum of probe electron (whose initial average momentum was zero) to be:
\begin{eqnarray}
p_{a0z}^{final}=-\frac{e^{2}TA_{s1}^{2}}{4\pi\epsilon_{0}}\cos\theta_{m}\int\limits_{b-\frac{a}{2}}^{b+\frac{a}{2}}dz\dfrac{\sin^{2}(k_{s1}(b+\frac{a}{2}-z))}{z^{2}}~~~~~~
\label{p_aoz,final}
\end{eqnarray}
In principle we can measure momentum of probe electron (and hence $\theta_m$) with arbitrary precision \footnote[9]{}. We note that, both $T\rightarrow\infty$ and separation between two potential wells tending to infinity are necessary. In the latter limit, the integral in eq. \ref{p_aoz,final} tends to zero (as integration is over only upper well), there by keeping the product, $T\times integral$, finite. We also note that, in an exact calculation (i.e., without doing any approximations), the functional form of eq. \ref{p_aoz,final} may change, but still it will be a function of $\theta_m$. In eq. \ref{p_aoz,final}, except $\cos\theta _{m}$ all other terms are greater than zero. If the unknown $\theta_{m}$ happens to be zero, there is no splitting of wave packet and the system electron will be in upper well for $B_i<0$, according to eq. \ref{splitting stage}. Hence, probe electron experiences maximum downward force(repulsion) and hence gains maximum negative momentum, which is consistent with  $p_{a0z}^{final}$ in eq. \ref{p_aoz,final}. If $\theta_{m}=\frac{\pi}{2}$, wave packet splits equally and hence probe electron experiences equal and opposite forces, gaining zero net momentum. $p_{a0z}^{final}$ in eq. \ref{p_aoz,final}, also gives same result. Finally if $\theta_{m}=\pi$, again no splitting of wave packet, but this time system electron will be in lower well. As a result, probe electron experiences maximum upward repulsive force, gaining maximum positive momentum. Again $p_{a0z}^{final}$ in eq. \ref{p_aoz,final} is consistent with it.

If we just want to discriminate between two nonorthogonal states, say $\ket{0}$ and $(\ket{0}+\ket{1})/\sqrt{2}$, then we have succeeded in discriminating, as it requires only the knowledge of polar angle $\theta_{m}$. However, if we want to clone the unknown state completely, we need to still findout its azimuthal angle, $\phi_{m}$ (\ref{unknown state}). As the state of system electron is unaltered at the end of protective measurement (\ref{final most state of sys+probe}), and \textit{because the trapping process was adiabatic} (see below eq. \ref{ket phi_dwp}), \textit{it must be reversible}, and hence we can get back the state in eq. \ref{splitting stage}, by adiabatically untrapping (procedure for untrapping is same as for trapping (below eq. \ref{ket phi_dwp}), but in reverse). Now we can recombine the split wave packets \footnote[6]{} by reversing the direction of inhomogeneous magnetic field which was used initially for splitting i.e., subjecting it to a field: $\vec{B}=B\hat{B}=-B_ix~\hat{i}+(-B_0+B_iz)\hat{k}$ (refer \footnote[1]{}), where, $B_i>0$ (note that while splitting, $B_i$ was $<0$). We recover the initial state, $\ket{\psi(0,p_{z})}=\ket{\hat{m}}\bar{\chi}(p_{z}-0)$ (\ref{splitting stage}), which had  $\Delta p_z\rightarrow 0$, but with $\Delta z\Delta p_z>\hbar/2$ \footnote[11]{It is not an issue, as increase in $\Delta z$ can be compensated by increasing width `a' of potential well, which can be arbitrarily large in principle.}. Polarization, $\hat{m}$, of unknown spin state is unaltered w.r.t lab frame. Previously we had applied inhomogeneous magnetic field along positive $z$-axis of lab frame. Now we apply it along a unit vector,
\begin{equation}
\hat{n}=\sin\theta_{n}\cos\phi_{n}\hat{i}+\sin\theta_{n}\sin\phi_{n}\hat{j}+\cos\theta_{n}\hat{k}
\label{n cap defined}
\end{equation}
where, $\theta_{n}\ne 0$ (zero is the value corresponding to previously applied direction of inhomogeneous magnetic field) and $\phi_{n}\ne 0$. From eq.s \ref{n cap defined} and \ref{m cap decomposed} we obtain the following constraint equation:
\begin{equation}
\hat{m}.\hat{n}=\cos\theta_{mn}=\cos\theta_{m}\cos\theta_{n}+\cos(\phi_{m}-\phi_{n})\sin\theta_{m}\sin\theta_{n}
\label{cos theta mn evalted}
\end{equation}
Component of unknown spin angular momentum operator along $\hat{n}$ is, $S_{n}=\vec{S}.\hat{n}$, where $\vec{S}=S_{x}\hat{i}+S_{y}\hat{j}+S_{z}\hat{k}=S_{m}\hat{m}$, $S_m$ is the component along itself. Eikenkets of operator $S_{n}$ are:
\begin{eqnarray}
\ket{+}_{n}=\cos\frac{\theta_{n}}{2}\ket{0}+\sin\frac{\theta_{n}}{2}e^{i\phi_{n}}\ket{1}\nonumber\\
\mathrm{and}~~\ket{-}_{n}=-\sin\frac{\theta_{n}}{2}\ket{0}+\cos\frac{\theta_{n}}{2}e^{i\phi_{n}}\ket{1}
\label{eigenkets of S_n}
\end{eqnarray} 
with eigenvalues $\frac{\hbar}{2}$ and $\frac{-\hbar}{2}$ respectively. In the eigenbasis of $S_{n}$, unknown spin state (\ref{unknown state}) can be decomposed as:
\begin{eqnarray}
\ket{\hat{m}}=\cos\frac{\theta_{mn}}{2}\ket{+}_{n}+\sin\frac{\theta_{mn}}{2}e^{i\phi_{mn}}\ket{-}_{n}
\label{unknown state in eign basis of S_n}
\end{eqnarray} 
where, $\theta_{mn}$ is given by eq. \ref{cos theta mn evalted} and $\phi_{mn}$ is the azimuthal angle of the unit vector $\hat{m}$, in a coordinate system rotated w.r.t the fixed lab frame, in which $\hat{n}$ is along it's positive $z$-axis. Inhomogeneous magnetic field along $\hat{n}$ is switched on for an interval of time $\tau$. System electron evolves under the interaction Hamiltonian $H_{int}=-\vec{\mu}.B_{i}q_n\hat{n}=-\gamma B_{i}q_nS_{n}$, where $q_n$ is the distance measured from origin along the direction $\hat{n}$. State of system electron after interacting with inhomogeneous magnetic field is given by:
\begin{eqnarray}
\ket{\psi_{\hat{n}}(\tau,p_n)}=
\cos\frac{\theta_{mn}}{2}\ket{+}_{n}\bar{\chi}(p_n-\gamma B_{i}\frac{\hbar}{2}\tau)\nonumber\\+\sin\frac{\theta_{mn}}{2}e^{i\phi_{mn}}\ket{-}_{n}\bar{\chi}(p_n+\gamma B_{i}\frac{\hbar}{2}\tau)
\label{splitting stage 2nd prot mesrmnt}
\end{eqnarray} 
where, $p_n$ is the component along $\hat{n}$ of momentum of system electron i.e., momentum conjugate to $q_n$. As before, trapping the system electron in a double well potential and measuring it protectively with a probe electron, we obtain the value of $\cos\theta_{mn}$. Hence we obtain the constraint eq. \ref{cos theta mn evalted} with the only unknown $\phi_{m}$. However we cannot obtain the value of $\phi_{m}$ unambiguously from a single constraint equation for the following reason: Consider the constraint eqation, $\cos\eta=c$, $-1\le c\le 1$. If $\eta$ takes a value in the interval $[0,\pi]$ (just like $\theta_{m}$), then there is only one value of $\eta$ which satisfies $\cos\eta=c$. On the other hand, if $\eta$ takes a value in the interval $[0,2\pi)$ (just like $\phi_{m}$), then there are two values of $\eta$ which satisfies $\cos\eta=c$. Hence we require one more constraint equation independent of the constraint eq. \ref{cos theta mn evalted}, to find out $\phi_{m}$ unambiguously. This is justified by the fact that, to unambiguously characterize an arbitrary orientation in space, we require three Euler angles, which are linearly independent of each other. Whereas, here we are trying to characterize the unknown direction, $\hat{m}$, with just two parameters $\theta_{m}$ and $\phi_{m}$, and hence the ambiguity. 

After obtaining the value of $\cos\theta_{mn}$, we shall again recombine the split wave packets, as before. Applying an inhomogeneous magnetic field along the direction $\hat{l}$ given by,
\begin{equation}
\hat{l}=\sin\theta_{l}\cos\phi_{l}\hat{i}+\sin\theta_{l}\sin\phi_{l}\hat{j}+\cos\theta_{l}\hat{k}
\label{l cap defined}
\end{equation}
where, $\theta_{l}\ne 0,\theta_{n}$ ($0,\theta_{n}$ are values corresponding to previously applied directions of inhomogeneous magnetic field) and $\phi_{l}\ne 0,\phi_{n}$. By this we obtain another independent constraint equation: 
\begin{equation}
\hat{m}.\hat{l}=\cos\theta_{ml}=\cos\theta_{m}\cos\theta_{l}+\cos(\phi_{m}-\phi_{l})\sin\theta_{m}\sin\theta_{l}
\label{cos theta ml evalted}
\end{equation}
Through protective measurement we can obtain the value of $\cos\theta_{ml}$. Solving eq.s \ref{cos theta mn evalted} and \ref{cos theta ml evalted} for $\phi_{m}$ and picking out the solution common to both of them, we obtain an unambiguous value of $\phi_{m}$. 

\section{Miscellaneous}
\textbf{Discriminating between $\ket{0}$ and $\ket{+}(=(\ket{0}+\ket{1})/\sqrt{2})$ even when the initial state is in a linear superposition of nondegenerate eigenstates of $H_{S}$ (\ref{H_S defined}):} Let the initial state of system and probe electron be: $\ket{\zeta'(0)}=\sum\limits_{i=1}^{\infty}c_{i}\ket{\phi''_{i}(z)}\bar{\Theta}(p_{az})$, where, $\ket{\phi''_{i}(z)}$ is defined in eq. \ref{ket phi'' defined}. Hence we need not worry about pushing the system electron into one of the eigenstates of $H_{S}$, unlike in cloning. Hence discrimination is more feasible than cloning. Allowing $\ket{\zeta'(0)}$ to evolve under the Hamiltonian $H$ given in eq. \ref{H defined}, for an interval of time $T$, and inserting completeness relations we obtain:
\begin{widetext}
\begin{eqnarray}
\ket{\zeta'(T)}=\sum\limits_{i=1}^{\infty}c_{i}\int\limits_{-\infty}^{\infty}dz_{a}\int\limits_{-\infty}^{\infty}dz'\sum\limits_{k=1}^{\infty}\exp\bigg(\frac{-i}{\hbar}\bigg(H_{S}+\frac{1}{4\pi\epsilon_{0}}\dfrac{e^2}{|\hat{z}-\hat{z}_{a}|}\bigg)T\bigg)\delta(z_{a})\outpr{\tilde{\phi}_{k}(z)}{\tilde{\phi}_{k}(z')}\ket{\phi''_{i}(z')}\bar{\Theta}(p_{az})
\label{ket zeta' comp insertd}
\end{eqnarray}
\end{widetext}
This is similar to the state in (\ref{ket zeta comp insertd}) but with an extra summation over the index `$i$'. As we are doing Protective Measurement, criteria of adiabaticity and weak interaction are satisfied and hence as before (below eq. \ref{eigenvale E tilde def}) we can take, $\ket{\tilde{\phi}_{k}(z)}\approxeq\ket{\phi''_{k}(z)}$. Proceeding in a fashion similar to that from eq. \ref{ket zeta comp insertd} to \ref{p_aoz,final}, we obtain,
\begin{eqnarray}
\ket{\zeta'(T)}=\sum\limits_{i=1}^{\infty}c_{i}~\exp\big(\frac{-i}{\hbar}\bigg(\expec{H_{S}}_{\phi''_{i}}+\frac{1}{4\pi\epsilon_{0}}e^{2}f_{i}(0)\bigg)T\big)\nonumber\\\ket{\phi''_{i}(z)}\bar{\Theta}\bigg(p_{az}+\frac{1}{4\pi\epsilon_{0}}e^{2}f'_{i}(0)T\bigg)~~~~~
\label{final most state of sys+probe zeta'(T)}
\end{eqnarray}
where,
\begin{eqnarray}
f'_{i}(0)=\cos\theta_{m}~A_{i}^{2}\int\limits_{b-\frac{a}{2}}^{b+\frac{a}{2}}dz~\dfrac{\sin^{2}(k_{i}(b+\frac{a}{2}-z))}{z^{2}}~~~~
\label{f'_i(0) evaluated}
\end{eqnarray}
If the given state is $\ket{+}$, then $\theta_{m}=\frac{\pi}{2}$ and hence $f'_{i}(0)=0$ for all `$i$'. As a result, there is no entanglement between system electron and probe electron (eq. \ref{final most state of sys+probe zeta'(T)}) and probe electron gains zero momentum. This makes sense, because, when $\theta_{m}=\frac{\pi}{2}$, wave packet of system electron spilts equally. As a result probe electron experiences equal and opposite repulsive forces, gaining no net momentum. However if the state is $\ket{0}$, then $\theta_{m}=0$ and hence $f'_{i}(0)$ is strictly greater than zero as $A_{i}$ being normalization and completeness satisfying constant, cannot be zero and $k_{i}$ (\ref{E reltd to k}) which corresponds to the energy of the $i^{th}$ level, cannot also be zero. Hence in this case, probe electron gains non-zero momentum (of course, in this case there will be collapse upon measurement as there is entanglement, but it does not matter as we are interested only in discrimination). Also, wave packet of system electron doesnot split and hence it will be in the upper well. As a result probe electron experiences downward repulsive force, there by gaining downward momentum. By measuring the momentum of probe electron (which can be measured as precisely as we want, at the cost of loosing information about its position, which is not required for discrimination, any way) we can \textit{always} discriminate between the states $\ket{0}$ and $\ket{+}$ unambiguously. However, we obtained eq. \ref{f'_i(0) evaluated} under the approximation given in expression \ref{H_int approx} and a few other approximations. Suppose there is a small correction $\Delta f'_{i}(0)$ to $f'_{i}(0)$ in eq. \ref{f'_i(0) evaluated}, under an exact calculation. By choosing $a$ and $b$ suitably, we can make the integral along with $A_{i}^{2}$ in eq. \ref{f'_i(0) evaluated} as large as we want. When $\theta_{m}=\frac{\pi}{2}$, $f'_{i}(0)=\Delta f'_{i}(0)$, where as when $\theta_{m}=0$, $f'_{i}(0)>>\Delta f'_{i}(0)$. Hence, we can still discriminate between $\ket{0}$ and $\ket{+}$ always. However, we note that, $\Delta f'_{i}(0)$ may also increase as the integral in eq. \ref{f'_i(0) evaluated} increases, in which case above argument breaks down. Also it may happen that under an exact calculation, there might be entanglement between system electron and probe electron even for $\theta_{m}=\frac{\pi}{2}$, in which case we cannot discriminate, as probe electron also gains non-zero momentum. One has to do exact calculation and see.

\textbf{How Not to Clone:} It's not possible to clone an arbitrary unknown spin state $\ket{\hat{m}}$ (\ref{unknown state}) by protecting it with a strong homogeneous magnetic field ($\vec{B'_{0}}$). As the direction, $\hat{m}$, of unknown spin polarization is not known a priori, we cannot apply $\vec{B'_{0}}$ along $\hat{m}$, in general. Hence we choose the direction of $\vec{B'_{0}}$ to be along $\hat{k}$ (positive $z$-axis of fixed lab frame) i.e., $\vec{B'_{0}}=B'_{0}\hat{k}$. Apply an inhomogeneous magnetic field $\frac{B_{i}}{T}q_{n}\hat{n}$, where, $q_{n}$ is the distance from origin measured along $\hat{n}$ (\ref{n cap defined}), $\frac{B_{i}}{T}$ is the field gradient in $Tesla~m^{-1}$ and in this context, $T$ is a dimensionless number which controls the gradient strength. As, $B'_0$ can be arbitrarily large, we can neglect $p_z^2/(2M)$ compared to $|-\vec{\mu}.B'_0\hat{k}|$. Hence, $\ket{\hat{m}}$ evolves under the Hamiltonian,
\begin{eqnarray}
H=-\vec{\mu}.B'_{0}\hat{k}-\vec{\mu}.B_{i}\frac{q_{n}}{T}\hat{n}=-\vec{\mu}.(B'_{0}\hat{k}+B_{i}\frac{q_{n}}{T}\hat{n})\nonumber\\=-\gamma\vec{S}.\vec{B}=-\gamma BS_{B}~~~~~~~~~~~~
\label{H static field}
\end{eqnarray}
where, $\vec{S}$ is the total spin angular momentum of unknown spin. $S_{B}=\vec{S}.\hat{B}$, where $\hat{B}$ is a unit vector along resultant magnetic field $\vec{B}$. In the eigenbasis of the operator $S_{B}$, state of unknown spin-$\frac{1}{2}$ particle at time $t=T$ is,
\begin{eqnarray}
\ket{M'(T)}=\cos\frac{\theta_{Bm}}{2}e^{-\frac{i}{2}\phi_{B}}\ket{+}_{B}\exp(-\frac{i}{\hbar}(-\gamma B)\frac{\hbar}{2}T)\bar{\chi}(p_n)\nonumber\\+\sin\frac{\theta_{Bm}}{2}e^{\frac{i}{2}\phi_{B}}\ket{-}_{B}\exp(\frac{i}{\hbar}(-\gamma B)\frac{\hbar}{2}T)\bar{\chi}(p_n)~~~~~~~~~~
\label{ket M(T)}
\end{eqnarray}
where $p_n$ is the componenet of momentum of unknown spin-$\frac{1}{2}$ particle along $\hat{n}$, $\cos\theta_{Bm}=\hat{B}.\hat{m}$. In the limit $T\rightarrow\infty$, which corresponds to adiabatic and weak interaction limit, we obtain: $B=|\vec{B}|\approxeq B'_{0}+\frac{B_{i}}{T}q_{n}\cos\theta_{n}$. But $\cos\theta_{n}=\frac{2}{\hbar}\expec{S_{n}}_{\ket{0}}$, where, $S_{n}=\vec{S}.\hat{n}$ and $\ket{0}$ is the eigenket of $S_{z}$. In this limit, state $\ket{M'(T)}$ becomes:
\begin{eqnarray}
\ket{M'(T\rightarrow\infty)}=~~~~~~~~~~~~~~\nonumber\\e^{-\frac{i}{2}(\phi_{B}-\gamma B'_{0}T)}\cos\frac{\theta_{Bm}}{2}\ket{+}_{B}\bar{\chi}\bigg(p_n-\gamma B_{i}\expec{S_{n}}_{\ket{0}}\bigg)~~~~~\nonumber\\+e^{\frac{i}{2}(\phi_{B}-\gamma B'_{0}T)}\sin\frac{\theta_{Bm}}{2}\ket{-}_{B}\bar{\chi}\bigg(p_n+\gamma B_{i}\expec{S_{n}}_{\ket{0}}\bigg)~~~~~~~~~~
\label{ket M(T), T going to infty}
\end{eqnarray}
In the limit $T\rightarrow\infty$, 
\begin{eqnarray}
\hat{B}.\hat{m}=\cos\theta_{Bm}\approxeq \cos\theta_{m}+\frac{B_{i}q_{n}}{B'_{0}T}(\cos\theta_{mn}-\cos\theta_{m}\cos\theta_{n})\nonumber\\
\label{B cap.m cap}
\end{eqnarray}
where, $\cos\theta_{mn}$ is given by eq. \ref{cos theta mn evalted}. Consider the case where, the direction of unknown spin polarisation, $\hat{m}$, happens to \textit{accidentally coincide} with that of static field direction, $\hat{k}$, and hence `rightly protected' (this is the case considered in \cite{prot_measure} as mentioned by themselves in \cite{mean_of_prot_mesrmnt}). In this case $\theta_{m}=0$ and we obtain $\cos\theta_{Bm}\approxeq 1$ $\Rightarrow \theta_{Bm}\approxeq 0$, $\hat{B}\approxeq\hat{m}=\hat{k}$, $\ket{+}_{B}\approxeq\ket{\hat{m}}=\ket{0}$ and the state \ref{ket M(T), T going to infty} becomes,
\begin{eqnarray}
\ket{M'(T\rightarrow\infty)}=e^{-\frac{i}{2}(\phi_{B}-\gamma B'_{0}T)}\ket{0}\bar{\chi}\bigg(p_n-\gamma B_{i}\expec{S_{n}}_{\ket{0}}\bigg)~~~~~~
\label{ket M(T), T going to infty accidentl coincidnce}
\end{eqnarray}
There is no entanglement in eq.\ref{ket M(T), T going to infty accidentl coincidnce} and the unknown spin state is unaltered. Upon measuring the momentum of spin-$\frac{1}{2}$ particle, we obtain:$\gamma B_{i}\expec{S_{n}}_{\ket{0}}=\gamma B_{i}\expec{S_{n}}_{\ket{\hat{m}}}$. Consider now the case where, the direction of unknown spin polarisation, $\hat{m}$, \textit{donot coincide}(which is highly highly probable) with that of static field direction, $\hat{k}$, and hence `wrongly protected'. In this case, $\cos\theta_{Bm}\ne 1$ as evident from (\ref{B cap.m cap}) (this is true even if $\theta_{mn}=0$ i.e., $\hat{m}$ accidentally coincides with $\hat{n}$ ) and hence spin and spatial d.o.f are entangled as given by eq. \ref{ket M(T), T going to infty}. Upon measuring the momentum of particle in momentum basis, we obtain either, momentum $\gamma B_{i}\expec{S_{n}}_{\ket{0}}$ and spin state collapses to $\ket{+}_{B}$ with probability $\cos^{2}\frac{\theta_{Bm}}{2}$, or momentum $-\gamma B_{i}\expec{S_{n}}_{\ket{0}}$ and spin state collapses to $\ket{-}_{B}$ with probability $\sin^{2}\frac{\theta_{Bm}}{2}$. When we obtain the outcome $\gamma B_{i}\expec{S_{n}}_{\ket{0}}$ there is no way to know if the spin state was rightly protected or wrongly protected, as it is common to both cases. However when we obtain $-\gamma B_{i}\expec{S_{n}}_{\ket{0}}$, we come to know that the spin state was wrongly protected. In either of the cases we obtain no information about the unknown spin state, $\ket{\hat{m}}$. Hence it's impossible to clone by this method.

\textbf{Cloning in Center of Wave Packet(CWP) Frame:} If we can somehow track the CWP of the split wave packets and measure its momentum, then we can perfectly clone an arbitrary unknown spin state, even with von-Neumann impulsive measurement, as follows: Consider the unknown state $\ket{\hat{m}}$ given in \ref{unknown state}. Switch on inhomogeneous magnetic field $B_{i}z\hat{k}$  for an interval of time $\tau$. Spin and sptial states evolve under the interaction Hamiltonian,
\begin{eqnarray}
H_{int}=-\vec{\mu}.B_{i}z\hat{k}=-\gamma \vec{S}.B_{i}z\hat{k}=-\gamma S_{m}\hat{m}.B_{i}z\hat{k}\nonumber\\=-\gamma B_{i}\cos\theta_{m}S_{m}z~~~~~~~~~~~~~~~
\label{H_int CWP frame}
\end{eqnarray}
where, $S_{m}=\vec{S}.\hat{m}$, and $S_{m}\ket{\hat{m}}=\frac{\hbar}{2}\ket{\hat{m}}$. We are working in the eigenbasis of the operator, $S_{m}$. State at time $t=\tau$ is,
\begin{eqnarray}
\ket{\psi(\tau,p_{z})}=\exp(\frac{i}{\hbar}\gamma B_{i}\cos\theta_{m}S_{m}z\tau)\ket{\hat{m}}\bar{\chi}(p_{z})\nonumber\\=\ket{\hat{m}}\bar{\chi}(p_{z}-\gamma B_{i}\cos\theta_{m}\frac{\hbar}{2}\tau)
\label{state at tau CWP frame}
\end{eqnarray}
Hence CWP has gained $z$-component of momentum equal to $\gamma B_{i}\cos\theta_{m}\frac{\hbar}{2}\tau$. If we can some how measure this momentum, we come to know $\theta_{m}$ and there is no collapse. By following a procedure similar to that described from eq.s \ref{n cap defined} to \ref{cos theta ml evalted}, we can find out $\phi_{m}$. However if we work in the eigenbasis of the operator $S_{z}$, we can observe the splitting. $H_{int}$ can be written in terms of $S_{z}$ as,
\begin{eqnarray}
H_{int}=-\vec{\mu}.B_{i}z\hat{k}=-\gamma \vec{S}.B_{i}z\hat{k}=-\gamma B_{i}S_{z}z\
\label{H_int not CWP frame}
\end{eqnarray}
and the state at time $t=\tau$ is exactly as given in eq.\ref{splitting stage}, where we can observe splitting and hence entanglement.

\section*{Conclusion}
In principle it is possible to do exact calculations without the approximations that we have made. Hence, we conclude that, in \textit{principle} it is possible to clone a single arbitrary unknown quantum state of a spin-1/2 particle (an electron) with arbitrary precision and with probability tending to one. Of course, in practice we cannot make dimensions of double well potential, $a$ and $b$, and the time interval $T$, arbitrarily large, even though possible in principle. Hence, it reduces the precision and success probability of cloning. However we note that, even a few millimeters and seconds are like infinity on the atomic scale. For eg., in NMR we talk of adiabatic processes even though life time of spins that we prepare is only a few seconds \cite{geophase_jones}. Hence, to get a quantitative picture of precision and success probability attainable in practice, one has to do thorough calculation considering all practical limitations. Our protocol can be easily generalized to arbitrary charged, spin-1/2 particle. If it is not electrically charged, then one can exploit its gravitational potential energy, in principle, as splitting of wave packet is independent of charge, but depends only on spin. Protocol can also be generalized to higher spin ($>1/2$) particles. Experimental realization of our protocol is with in the reach of current technology. 

Non-orthogonal state discrimination is more feasible than cloning, as it requires only the knowledge of polar angle, $\theta_m$. Also, to discriminate between, say, $\ket{0}$ and $\ket{+} (=(\ket{0}+\ket{1})/\sqrt{2})$, it is not necessary to find $\theta_m$ with arbitrary precision. So much error is allowed such that we can just unambiguously discriminate between $\ket{0}$ and $\ket{+}$. Here we note that, inspite of all practical constraints on $a,~b$ etc., if we can still unambiguously discriminate between $\ket{0}$ and $\ket{+}$ in finite time $T$ (such that $T$ is less than time interval corresponding to two space-like separated events) with success probability greater than, say, $0.5$, then our protocol may open up the doors to superluminal communication. To check this possibility, one has to do thorough analysis considering all practical constraints. We also note that, the limit $T\rightarrow\infty$ corresponds to discriminating between $\ket{0}$ and $\ket{+}$ with success probability tending to one, and it is not possible to do superlumial communication in this limit, as $T$ is greater than any finite time interval corresponding to two space-like separated events. But to communicate superluminally, we donot require success probability tending to one. Hence, there seems to be a possibility still.


\section*{Acknowledgement}
I am grateful to my guide Dr. T S Mahesh for proposing this idea and giving freedom to work on it. I am also thankful to him for many motivating discussions and ideas. I am also thankful to Prof. N D Hari Dass for introducing to protective measurements. I am grateful to Prof. Masanao Ozawa for valuable comments. I thank Govind Unnikrishnan for pointing to linearity.
\bibliographystyle{apsrev4-1}
\bibliography{bib_ch}

\end{document}